\documentclass[twocolumn,twocolappendix]{openjournal}

\usepackage[utf8]{inputenc}
\usepackage[T1]{fontenc}
\usepackage{float}
\usepackage{ulem}
\usepackage{graphicx}	% Including figure files
\usepackage{amsmath}	% Advanced maths commands
\usepackage{longtable}
\usepackage{comment}
\usepackage[dvipsnames,svgnames]{xcolor}
\usepackage{color}
\usepackage{hyperref}
\hypersetup{colorlinks=true,
	urlcolor=blue,  % color of external links
	linkcolor=blue,  % color of toc, list of figs etc.
	citecolor=blue, % color of links to bibliography
    menucolor=blue, % color for Acrobat menu buttons
    urlcolor=blue}  % color for \url{...} links

\def\msun{\, M_{\odot}}
\def\lsun{\, L_{\odot}}

\def\simlt{\lower.5ex\hbox{$\; \buildrel < \over \sim \;$}}
\def\simgt{\lower.5ex\hbox{$\; \buildrel > \over \sim \;$}}

\def\contourtext{The contours contain $1\sigma$ and $2\sigma$ of galaxies in their subpopulation.}
\newcommand{\astrid}{\texttt{ASTRID}}
\newcommand{\brahma}{\texttt{BRAHMA}}

\def\LRDnum{41}
\def\notCompactLRDnum{57}

\def\RedOneLRDnum{37}
\def\RedTwoLRDnum{4}

\def\contourtext{The magenta contours contain $1\sigma$ and $3\sigma$ of sources in the \astrid\ simulation}
\def\RedColorText{The 2D histogram shows the overall \brahma\ population. It is colored by the mean ``Rest-visible Color'' in each cell, where ``Rest-visible Color'' is F200W-F356W for z=5,6 sources, and F277W-F444W for z=7,8 sources}

\newcommand{\rev}[1]{#1}

\begin{document}
% Title of the paper, and the short title which is used in the headers.
% Keep the title short and informative.
\title{From ASTRID to BRAHMA -- The role of overmassive black holes in little red dots in cosmological simulations}
% The list of authors, and the short list which is used in the headers.
\author{Patrick LaChance$^{1,2,*}$}
\author{Aklant Kumar Bhowmick$^{3,4,5}$}
\author{Rupert A.C. Croft$^{1}$}
\author{Tiziana Di Matteo$^{1}$}
\author{Yihao Zhou$^{1}$}
\author{Fabio Pacucci$^{6,7}$}
\author{Laura Blecha$^{8}$}
\author{Paul Torrey$^{3,4,5}$}
\author{Yueying Ni$^{9}$}
\author{Nianyi Chen$^{10}$}
\author{Simeon Bird$^{11}$}
\thanks{$^*$E-mail: P.LaChance@soton.ac.uk}
% List of institutions
\affiliation{$^{1}$ McWilliams Center for Cosmology and Astrophysics, Department of Physics, Carnegie Mellon University, Pittsburgh, PA 15213 USA } %\\}
\affiliation{$^{2}$ School of Physics \& Astronomy, University of Southampton, Highfield, Southampton SO17 1BJ, UK} %\\}
\affiliation{$^{3}$ University of Virginia 530 McCormick Rd Charlottesville, VA 22904, USA } %\\}
\affiliation{$^{4}$ Virginia Institute for Theoretical Astronomy, University of Virginia, Charlottesville, VA 22904, USA } %\\}
\affiliation{$^{5}$ The NSF-Simons AI Institute for Cosmic Origins, USA } %\\}
\affiliation{$^{6}$ Center for Astrophysics $\vert$ Harvard \& Smithsonian, 60 Garden St, Cambridge, MA 02138, USA} %\\}
\affiliation{$^{7}$ Black Hole Initiative, Harvard University, 20 Garden St, Cambridge, MA 02138, USA} %\\}
\affiliation{$^{8}$ Department of Physics, University of Florida, Gainesville, FL 32611, USA} %\\}
\affiliation{$^{9}$ Michigan Institute for Data and AI in Society,
500 Church Street, Suite 600, Ann Arbor, MI 48109} %\\}
\affiliation{$^{10}$ Institute for Advanced Study, 1 Einstein Dr, Princeton, NJ 08540} %\\}
\affiliation{$^{11}$ Department of Physics \& Astronomy, University of California, Riverside, 900 University Ave., Riverside, CA 92521, USA} %\\}

% Abstract of the paper
\begin{abstract}
We investigate the origin of Little Red Dots (LRDs) using a realization of the \brahma\ cosmological hydrodynamic simulation that produces overmassive black holes at $z \sim 4\text{-}8$. We generate mock \textit{JWST} observations by modeling galaxy spectral energy distributions (including dust attenuation) together with AGN emission using \textsc{cloudy}-based templates, and test whether such systems reproduce the observed LRD population. 
\rev{We consider both a standard low-redshift AGN emission model and a gas-enshrouded AGN emission model, in which the AGN is embedded within unresolved dense gas that produces a strong Balmer break, and find that LRDs emerge in significantly larger numbers with the latter model.}
Assuming that all AGN reside in this gas-enshrouded phase, we recover LRD number densities ($\Phi \sim 2 \times 10^{-4}~\mathrm{Mpc}^{-3}$), close to the observed abundance. This implies an effective gas-enshrouded duty cycle approaching unity, which is not predicted by current theoretical models. Current JWST constraints on the high-z black hole mass function suggest that this tension may arise from a deficit of low-mass ($M_{\rm BH}\lesssim10^7~M_\odot$) LRDs in \brahma. Applying the same gas-enshrouded model to the \astrid\ simulation, which lacks overmassive black holes, we find that the LRD abundance remains $\gtrsim2~\rm dex$ below observations. Overall, while the overmassive black holes in this \brahma\ realization substantially enhance the production of LRD-like systems, reproducing the observed LRD abundance requires them to reside in dense gas environments with unrealistically high duty cycles. We therefore conclude that overmassive black holes are unlikely to be the sole origin of the observed JWST LRD population, and may instead account for only a subset of these systems.
\end{abstract}

% Select between one and six entries from the list of approved keywords.
% Don't make up new ones.

\maketitle

%%%%%%%%%%%%%%%%%%%%%%%%%%%%%%%%%%%%%%%%%%%%%%%%%%
%%%%%%%%%%%%%%%%% BODY OF PAPER %%%%%%%%%%%%%%%%%%

\section{Introduction}
\label{sec:Intro}

Early JWST observations revealed a population of objects dubbed ``little red dots'' \citep[LRDs;][]{Matthee_2023, Labbe_2023, Kocevski_2023, Harikane_2023, Furtak_2023}. These objects are compact in appearance, and feature a very red color in the rest-frame optical bands, and a flat or blue color in the rest-frame UV bands. Since their discovery, it has been unclear what they are, with initial proposals including compact star-forming galaxies and dust-attenuated AGN \citep{Baggen_2023, Barro_2023, Labbe_2023, Maiolino_2024_JADES, Guia_2024}.

A variety of follow-up observations have been conducted on these objects, including JWST spectroscopy \citep{Kokorev_2024}, X-ray \citep{Ananna_2024, Yue_2024, Maiolino_2025}, and infrared observations \citep{Casey_2025}, which have identified a number of common features of the spectra of LRDs. They are X-ray weak \citep{Akins_2025, Ananna_2024, Yue_2024, Maiolino_2025, Pacucci_Narayan_2024}, feature broad Balmer emission lines \citep{Wang_2024, Ma_2024}, and a significant Balmer break \citep{Setton_2024, Labbe_2024}, and there are upper limits on their infrared \citep{Casey_2025} and radio \citep{Perger_2025} emission. This combination of features, along with their UV and visible colors and compact morphology, has made them difficult to reconcile with existing models of galaxy formation, stellar populations, and AGN emission. Dust-attenuated sources, both stellar and AGN, would likely be far brighter in the IR than observed LRDs due to dust reprocessing \citep{Casey_2024}. Traditional AGN emission models without significant dust attenuation would not produce the visible red colors of observed LRDs, and may produce more X-ray emission than is expected from current observations \citep{Shen_2020}. Dense stellar sources would generally lack the broad emission lines seen in LRD spectra, but may be able to produce them in specific circumstances \citep{Perez_Gonzalez_2024, Hviding_2025, Baggen_2024, Guia_2024, Pacucci_Hernquist_2025}.

Recent results have placed an increasing amount of importance on the Balmer break as a characteristic feature of the LRD spectrum \citep{Liu_2025}, and a variety of works have proposed AGN-centric models which feature spectral energy densities (SEDs) that differ significantly from those that are based on catalogs of low redshift quasars \citep{Shen_2020}. Two approaches have been taken to this. One derives an adjustment to the bolometric corrections for AGN using a purely empirical approach based on the observations of two of the brightest observed LRDs, and extrapolate a possible AGN SED from the observed emission, and detection limits \citep{Greene_2025}. The other utilizes photoionization modeling to produce AGN spectra, and attempt to determine the AGN parameters that fit observed LRD spectra. This process has been applied to CAPERS-LRD-z9 \citet{Taylor_2025} MoM-BH*-1 \citet{Naidu_2025}, and \textit{The Cliff} \citep{deGraff_2025}, and all three cases find atypical properties for the AGN and surrounding gas.

Specifically, they find that enshrouding the AGN within a very dense cloud of gas, which produces a pronounced Balmer break, results in spectra that can reproduce many of the observed properties of LRDs. The Balmer break creates the red color in the rest optical region, and the bright AGN produces the broad emission lines in that wavelength region. The dense gas environment would also limit the amount of X-ray emission that could escape the AGN, which aligns with the observations that find the vast majority of LRDs are X-ray weak. Additionally, the reddening provided by the gaseous environment limits the amount of dust attenuation that is required to appear as an LRD, which also limits the dust re-emission in the infrared, aligning with the observed upper limits on LRD infrared emission. Further motivation for these models is provided in \citet{Inayoshi_2025}, which performs an in depth analysis of the spectral features of ``gas-enshrouded'' AGN, and in \citet{Jeon_2025, Cenci_2025} which posit direct collapse black holes (DCBHs) as progenitors of the AGN in LRDs. \cite{Pacucci_2026_DCBH} also recently demonstrated that the observed properties of LRDs can be reproduced by assuming they are recently formed DCBHs enshrouded by gas.

The study of LRDs is also closely intertwined with investigations of the population of potentially ``overmassive'' black holes observed at similarly early epochs, some of which have been identified as LRDs \citep{Pacucci_2023, Durodola_2024, Taylor_2024, Furtak_2024, Jones_2025}. While there is the possibility that the masses of these objects have been overestimated \citep{Greene_2025, Brooks_2024, Brooks_2025}, it remains unclear which interpretation is correct. Further exploration of the physical conditions capable of producing such overmassive black holes—and whether these conditions also give rise to LRD populations consistent with observations—may help resolve this ambiguity.

Cosmological simulations may provide unique insights into these objects, as they couple large scale gravity+hydrodynamics with a wide range of relevant astrophysical processes such as cooling, star formation, black hole seeding, accretion, and feedback, to produce statistically large populations of galaxies and BHs. Given the breadth of possibilities for the nature of LRDs and the remaining uncertainties about their core physical properties, such as black hole and stellar mass, identifying and analyzing LRDs in simulations could provide clarity on the properties and environments necessary for their production. Most importantly, cosmological simulations offer the volume necessary to perform statistical analyses of LRDs to compare directly with observational constraints. This includes population level properties of LRDs such as their number density, and clustering.

Despite their advantages, cosmological simulations do not perform all of the astrophysical processes that are necessary for the modeling of LRDs. This includes many aspects crucial to modeling observations such as the production of emission from AGN and stars, and the attenuation of that emission by dust. As such, post-processing is required to produce observables from the simulation output. For the AGN this is primarily done via radiative transfer modeling with software like \texttt{CLOUDY} \citep{cloudy}, which is coupled to properties of the black holes from the simulation and produces the spectra of those black holes. As the radiative transfer occurs on length scales that are well below the resolution of the cosmological simulation, there are aspects of the model, such as the properties of the gas cloud surrounding the AGN, which cannot be inferred from the simulation. Similarly, the stellar emission is based on the properties of the star particles present in the simulation, via a coupling to a stellar synthesis model, as each star particle in the simulation represents many individual stars.

We previously performed an initial analysis of the LRD population in the \astrid\ simulation \citep{astrid_BHs, astrid_galaxy_formation, Ni_2025, Zhou_2025} and identified only a very limited number of LRD analogs, corresponding to a small fraction of the observed number density \citep{LaChance_2025}. There are several possible explanations for this deficit. First, the black holes in \astrid\ are substantially less massive, at fixed host-galaxy stellar mass, than the overmassive black holes inferred from recent JWST observations. Second, if LRDs are powered by AGN embedded within dense gaseous environments, then standard AGN SED models calibrated on lower-redshift AGN populations may not provide an adequate description of their emission. To assess the importance of these two effects, here we analyze another simulation selected for its unusually overmassive black hole population, while modeling its AGN using both standard and gas-enshrouded AGN SED prescriptions.

The recent \brahma\ simulations~\citep{Bhowmick_2024,Bhowmick_2024c} have systematically explored a wide range of black-hole seed model variations to identify the conditions necessary for producing overmassive black holes at high redshift. In particular, \cite{Bhowmick_2024c} demonstrated that overmassive black holes at $z \sim 4$–$7$ can be assembled if heavy $\sim10^{5}~M_{\odot}$ seeds form in sufficiently abundant numbers and subsequently grow efficiently through black-hole mergers occurring within $\lesssim750~\mathrm{Myr}$ following their host-galaxy mergers. These conditions may be optimistic, given the stringent requirements for direct-collapse black-hole formation and the possibility of longer delay times for black-hole mergers~\citep{2021MNRAS.508.1973M,2024MNRAS.532.4681P}. Nevertheless, the \brahma\ simulations provide an ideal arena for isolating and understanding the influence of black-hole ``overmassiveness'' on the occurrence of LRDs in cosmological simulations. Accordingly, in this work we focus on the \brahma\ simulation volume that produces the most overmassive black holes—consistent with current JWST constraints—by adopting a deliberately lenient model for heavy $\sim10^{5}~M_{\odot}$ seed formation, and we search for LRD-like objects in mock observations of the resulting galaxy and AGN populations at $z \sim 4$–$8$.

This paper is organized as follows. In Section~\ref{sec:methods}, we describe the simulations used in this work and the methodology employed to generate mock observations. In Section~\ref{sec:results}, we identify LRD analogs in \brahma\ using observational selection criteria and characterize their abundances, as well as their host-galaxy and black-hole properties. Finally, we discuss possible implications of our findings in Section~\ref{sec:discussion} and summarize our conclusions in Section~\ref{sec:Summary}.

\section{Methods}
\label{sec:methods}

%We produce mock observations of galaxies and AGN in a realization of the \brahma\ simulation suite, and analyze their properties, with a particular focus on identifying and analyzing the ``little red dot'' sources. We create our mock observations following the same process used in \citet{LaChance_2025} \aklant{\bf A bit redundant to the last part of intro and what follows. Could consider removing this}

\subsection{The \brahma\ Simulation Suite}
\label{subsec:brahma}

The \texttt{BRAHMA} cosmological simulation suite~\citep{Bhowmick_2024,Bhowmick_2024c} was run with the \texttt{AREPO} cosmological magnetohydrodynamics~(MHD) code \citep{2010MNRAS.401..791S, 2011MNRAS.418.1392P, 2016MNRAS.462.2603P, 2020ApJS..248...32W}, which solves gravity using a PM–Tree algorithm \citep{1986Natur.324..446B} and evolves the gas using the ideal MHD equations on a dynamically moving Voronoi mesh. The simulations adopt the \cite{2016A&A...594A..13P} cosmology and are initialized at $z=127$ using \texttt{MUSIC} \citep{2011MNRAS.415.2101H}. Many components of the galaxy formation model follow the \texttt{IllustrisTNG} framework \citep{Springel_2018, 2018MNRAS.475..648P, Nelson_2018, Naiman_2018, Marinacci_2018, 2019ComAC...6....2N}, including radiative cooling from primordial and metal species, star formation in gas above the density threshold of $0.13~\mathrm{cm}^{-3}$, stellar evolution and chemical enrichment, and stellar feedback implemented via hydrodynamically decoupled galactic winds. Black hole accretion follows an Eddington-limited Bondi prescription with a radiative efficiency of $\eta = 0.2$ %$\epsilon_r = 0.2$, 
while AGN feedback is modeled using the two-mode thermal/kinetic scheme employed in \texttt{IllustrisTNG}. For black holes, the \texttt{BRAHMA} suite contains simulations that employ both repositioning and subgrid dynamical friction. The specific simulation analyzed here adopts the former: black holes are repositioned toward the local gravitational potential minimum within a region defined by the 1000 nearest gas neighbors, and mergers are triggered when multiple BHs reside within this region.

A key feature of the \texttt{BRAHMA} simulations is their use of physically motivated BH seeding prescriptions~\citep{2021MNRAS.507.2012B,2022MNRAS.510..177B,2024MNRAS.529.3768B}, which are designed to capture the conditions expected in popular seed formation channels, such as Pop III remnants \citep{Fryer_2001, Madau_2001, Xu_2013, Smith_2018} and Direct Collapse Black Holes (DCBHs) \citep{Bromm_2003, Begelman_2006, Regan_2014, Latif_2016, Luo_2018, Luo_2020, Wise_2019, Begelman_2023, Mayer_2024}. This is achieved through a variety of seeding criteria that identify environments commonly associated with most seed formation channels, namely high gas densities and low gas metallicities, along with additional criteria that are particularly relevant for DCBH formation, such as sufficiently strong Lyman--Werner radiation fields, low gas angular momentum (or spin), and rich halo environments. The full \brahma\ suite incorporates a wide range of boxes with systematic variations in seeding criteria, focusing on lower-mass $\sim10^3~M_{\odot}$ seeds~\citep{Bhowmick_2024,2025ApJ...991...81B} as well as heavier $\sim10^4$--$10^5~M_{\odot}$ seeds~\citep{Bhowmick_2024c,Bhowmick_2025,bhowmick_2026}. As mentioned earlier, \cite{Bhowmick_2024c} demonstrated that the intrinsically overmassive high-$z$ BH populations are most readily reproduced in \brahma\ volumes that adopt the most lenient heavy-seed formation prescriptions.

Since our goal is to specifically investigate the impact of overmassive BHs on LRD populations, in this work we focus on a heavy-seed \brahma\ box that adopts the most lenient seeding criteria, requiring only high gas densities and low gas metallicities for seed formation. This realization does not include the Lyman–Werner radiation, angular momentum, or halo environment criteria that are employed in other simulations within the \brahma\ suite. Specifically, seed black holes are planted in halos that host a reservoir of \textit{dense} ($n_{\rm H}>0.13~\mathrm{cm}^{-3}$) and \textit{metal-poor} ($Z<10^{-4}Z_{\odot}$) gas with a mass equal to five times the black hole seed mass ($7.5\times10^{5}~M_{\odot}$). \cite{Bhowmick_2024c} applied this seeding model (labelled SM5) to an $[18~\mathrm{Mpc}]^{3}$ box (with $512^{3}$ DM particles) and demonstrated that the resulting BH population is systematically overmassive at $z\sim4$--7, in broad agreement with many of the overmassive BHs inferred from \textit{JWST}~\cite{Pacucci_2023, Juodžbalis_2025}. Here, we apply the same prescription to a larger $[36~\mathrm{Mpc}]^{3}$ box (with $1024^{3}$ DM particles) to investigate whether, and under what physical conditions, these overmassive BH populations may manifest observationally as the LRDs recently discovered by \textit{JWST}.

From a physical standpoint, although our seed masses are $\sim10^5~\msun$, the seeding criteria adopted in the selected \brahma\ box are too lenient to be interpreted as a direct model of the DCBH channel, particularly because they do not include a Lyman--Werner flux criterion. However, several alternative pathways could also produce $\sim10^5~\msun$ black hole seeds without requiring strong Lyman--Werner radiation. These include runaway stellar collisions in ultra-dense nuclear star clusters~\citep{2023PhRvD.108h3012K}, as well as episodes of hyper-Eddington accretion that can rapidly grow light seeds into the heavy-seed mass regime~\citep{2024OJAp....7E.107M,2026NatAs..10..583M}.

\subsection{The \astrid\ Simulation}
\label{subsec:astrid}

In addition to the \brahma\ simulation, we use the \astrid\ simulation \citep{astrid_BHs, astrid_galaxy_formation, Ni_2025, Zhou_2025} as a point of comparison throughout this work, as we have performed similar analysis of this simulation, and it contains black holes consistent with pre-JWST observations. It was run with a standard set of cosmological parameters (based on the Planck Survey \citet{Planck_2020}) and astrophysical models using the Smoothed Particle Hydrodynamics (SPH) code \texttt{MP-GADGET} \citep{Feng2018_MPGadget}. The simulation volume is $[\rm 369 \space Mpc]^3$, with an initial particle load of $\rm 2 \times 5500^{3}$ particles. The resulting mass resolution is $\sim6.5$ times lower than the \brahma~box we use, and similar to the Illustris TNG100 and EAGLE simulations, with a simulation volume larger than Illustris TNG300 (Illustris TNG: \citealt{Springel_2018, Nelson_2018, Marinacci_2018, Naiman_2018}; EAGLE: \citealt{Schaye_2015}).

The star formation is performed with the model described in \citet{Feng_2016}, which is based on the model originally developed in \citet{Springel_2003}. There is an additional correction to handle the formation of molecular hydrogen, and corresponding adjustments to the star formation model in low metallicity environments per \citet{Krumholz_2011}. The gas cooling follows the prescription of \citet{Katz_1996}, with an implementation of dense gas self-shielding per \citet{Rahmati_2013}. The full description of the stellar and gas processes are detailed in \citet{astrid_galaxy_formation} and \citet{astrid_BHs}.

Similar to \brahma, \astrid\ employs a modified Bondi--Hoyle accretion prescription. \astrid\ modifies the Bondi-Hoyle model by adding a boost factor of 100. It also differs from \brahma\ by using a lower radiative efficiency of 0.1, and allowing for mildly super-Eddington accretion, up to twice the Eddington limit. 
The AGN feedback model in \astrid\ is broadly similar to that in \brahma, with both thermal and kinetic feedback modes present. Despite a relatively lenient accretion model compared to \brahma, black hole growth in \astrid—particularly within low-mass galaxies—is significantly slower than in \brahma. The primary driver of this difference is the black hole seeding procedure. In \astrid, halos are seeded with initial black holes of mass $\sim4.4\times10^{4}$–$4.4\times10^{5}\,M_{\odot}$ only once both the total halo mass and stellar mass exceed thresholds of $\geq7.4\times10^{9}\,M_{\odot}$ and $3\times10^{6}\,M_{\odot}$, respectively. This criterion is substantially more restrictive than that adopted in the \brahma\ volume used here, where black hole seeds can form in halos with masses as low as $\sim10^{8}\,M_{\odot}$. Furthermore, \astrid\ incorporates a subgrid dynamical friction model to capture black hole dynamics \citep{Chen_2022}, whereas the particular \brahma\ box analyzed in this work employs black hole repositioning.

Together, these differences lead to more efficient merger-driven black hole growth in low-mass galaxies in \brahma, a behavior that is largely absent in \astrid. While the repositioning scheme is likely to overestimate merger-driven growth in \brahma, the substantially higher seeding frequency alone implies that \brahma\ should generically experience a larger number of black hole mergers than \astrid, even when coupled with a subgrid dynamical friction model. This efficient black hole growth in \brahma\ produces many differences between the black hole populations of \brahma\ and \astrid. This includes \brahma\ having $\sim 2$ dex more black holes at any given mass than \astrid\ in the redshift range we analyze. A more in-depth analysis of the impact of this seeding model on the black hole population is performed in \citet{AMBRA}. In this work we investigate the implications of the overmassive black hole population in \brahma\ on the formation and properties of LRDs, using our previous \astrid\ results as our baseline for comparison.

\begin{figure}
    \centering
    \includegraphics[width=1.0\columnwidth]{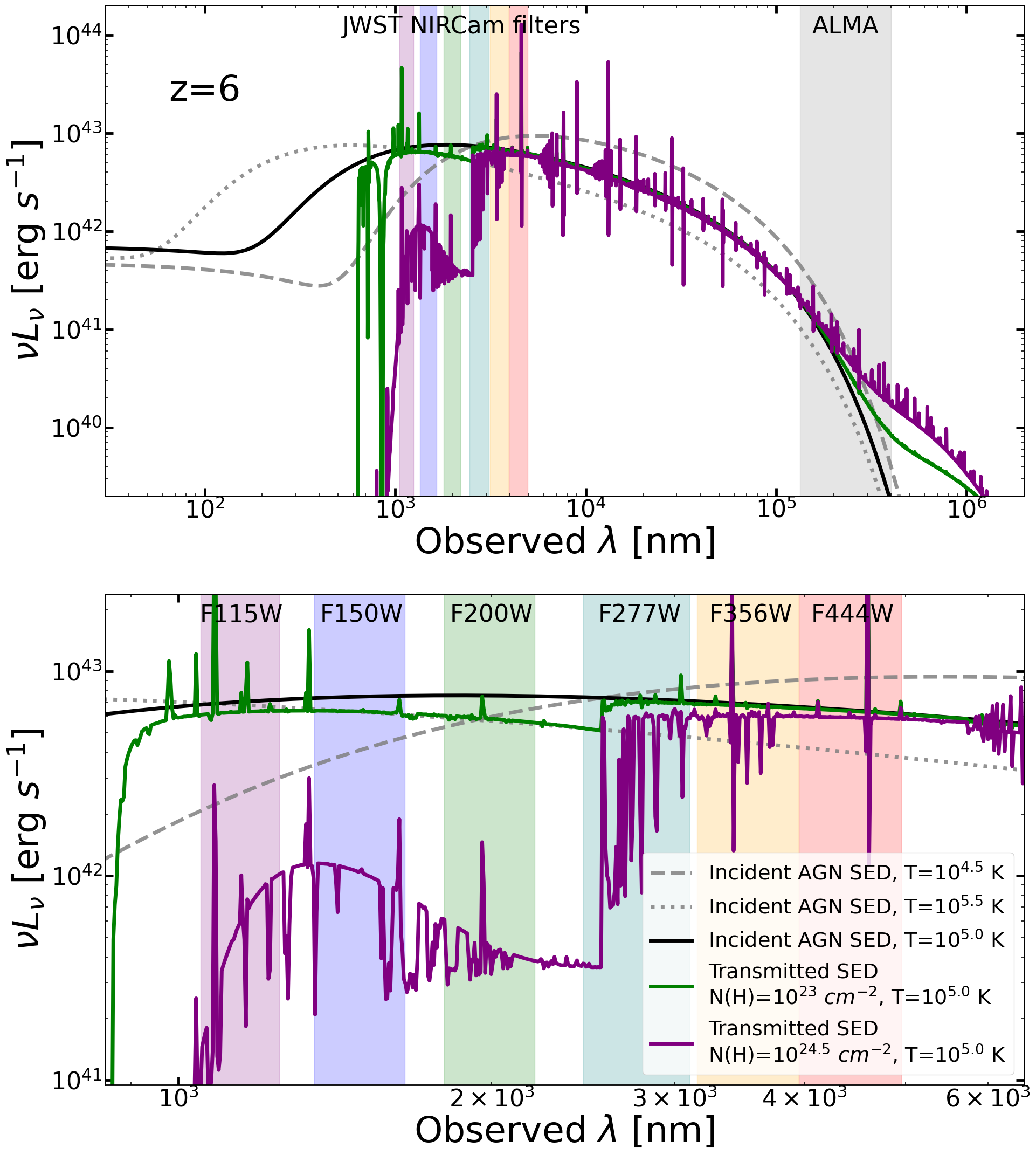}
    \caption{Example of the AGN Spectra used as input for, and output by, \texttt{CLOUDY}. We are placing this example object at $\rm z=6$ to provide relevant comparisons with observational bands. The top panel shows a broad view of the spectrum, spanning the infrared and X-ray regions. The lower panel zooms in on the JWST NIRCam wavelength range. The black line is the incident AGN spectrum in \texttt{CLOUDY}, which is processed through the surrounding gas cloud. We have chosen a temperature of $\rm T=10^{5.0} K$ for the example AGN. The green, and purple lines show different transmitted spectra based on the properties of the gas cloud. The green line is the cloud properties used in \citet{LaChance_2025}, and the purple spectrum is produced by the gas-enshrouded model we use in this work. The dashed and dotted grey lines are the incident AGN spectra produced by AGN with temperatures $\rm T=10^{4.5} K$, and $\rm T=10^{5.5} K$ respectively. At this bolometric luminosity ($10^{43.5} \rm erg/s$) the $\rm T=10^{4.5}$, $10^{5.0}$, $10^{5.5}$  K lines correspond to black holes with masses of $\rm \sim 10^{6}$, $10^{7}$, $10^{8} \msun$ respectively.}
    \label{fig:example_spectrum}
\end{figure}

\begin{figure}
    \centering
    \includegraphics[width=1.0\columnwidth]{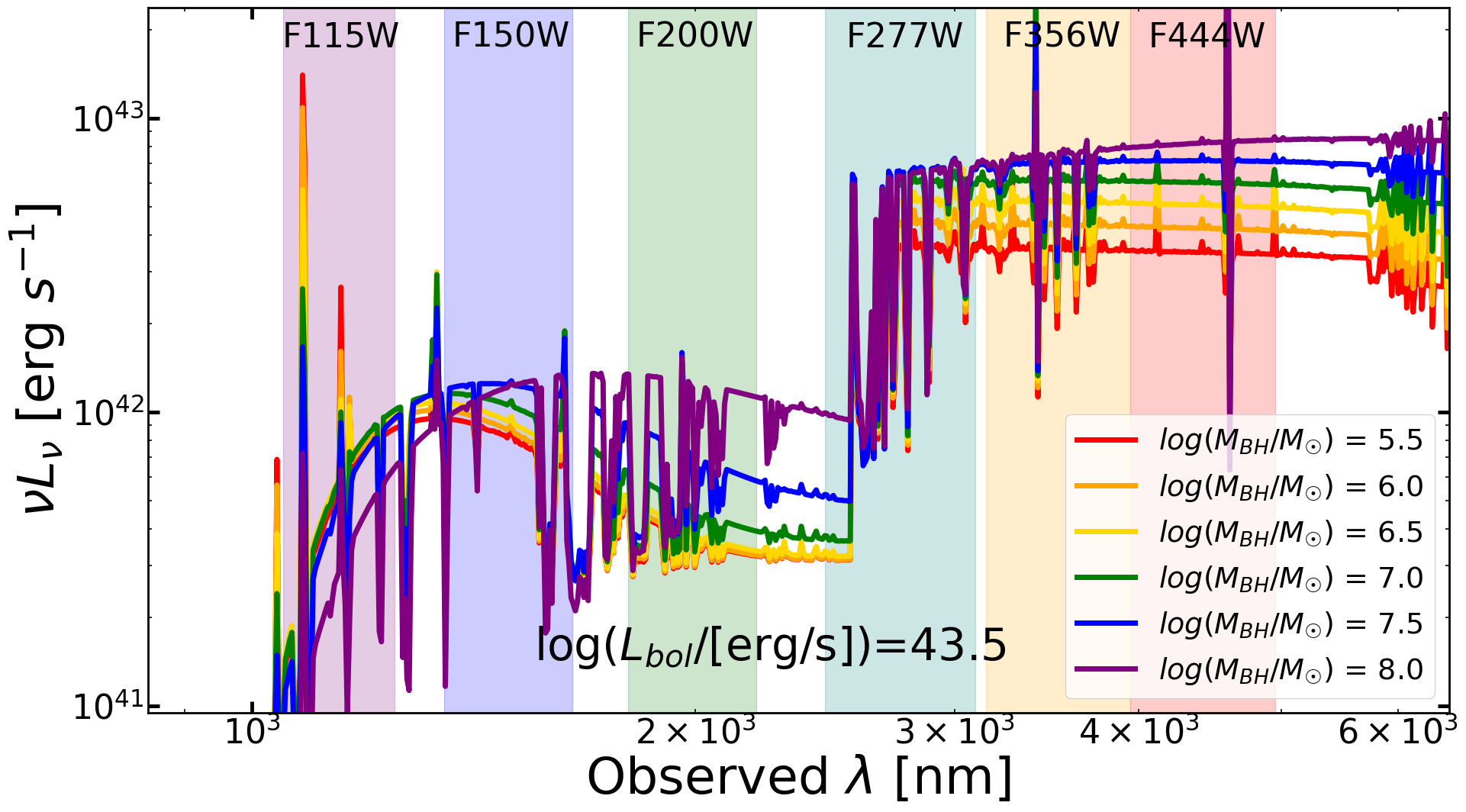}
    \includegraphics[width=1.0\columnwidth]{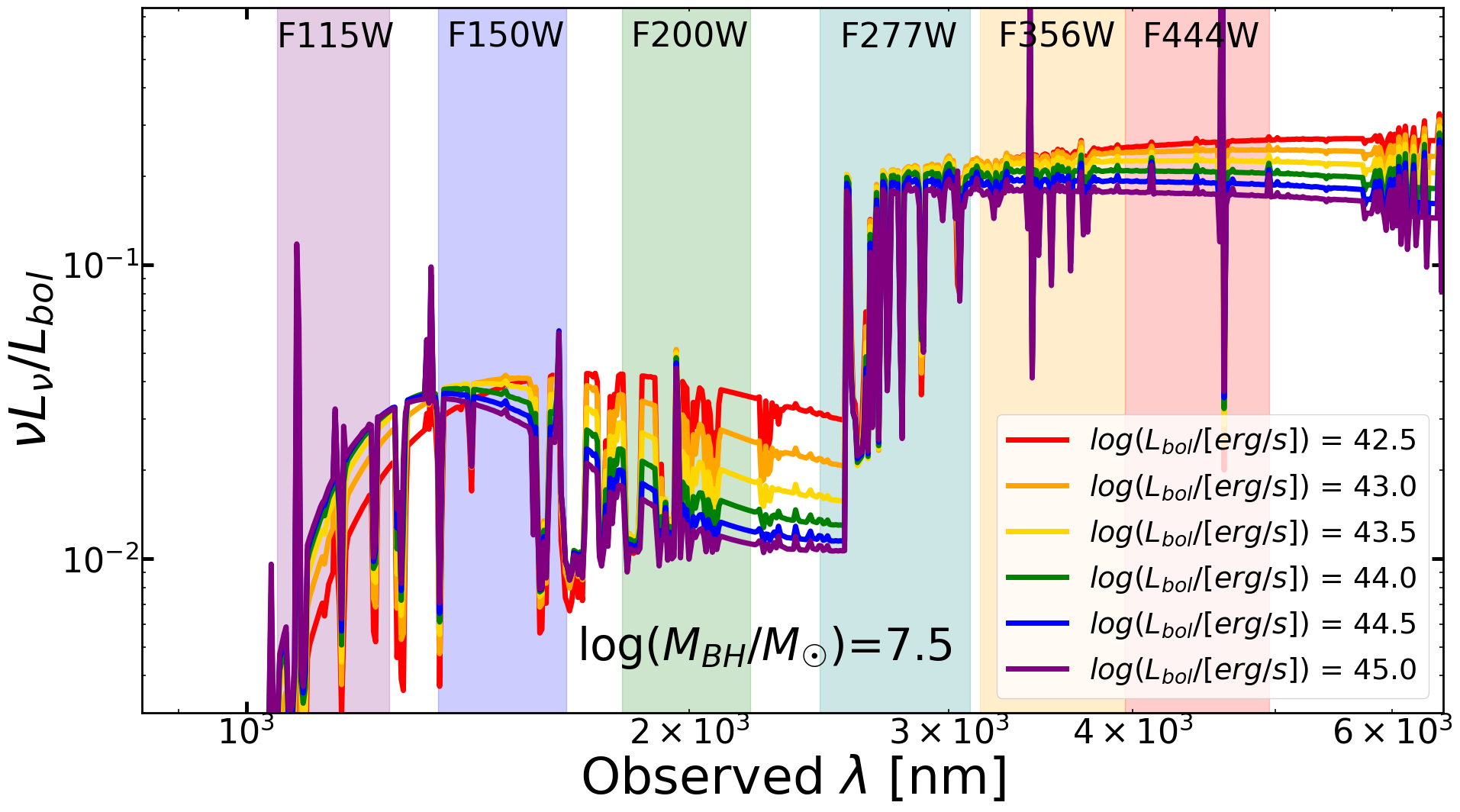}
    \caption{Examples of gas enshrouded AGN Spectra produced by, \texttt{CLOUDY} with varying black hole mass (upper panel) and varying bolometric luminosity (lower panel). We are treating these example spectra as though they are observed at $\rm z=6$ to provide relevant comparisons with observational bands. The AGN SEDs in the upper panel are all produced with a bolometric luminosity of $\rm 10^{43.5} erg/s$, and black hole masses ranging from $\rm M_{BH} = 10^{5.5} \msun$ to $10^{8.0} \msun$, which corresponds to a characteristic temperature range of $10^{5.7}$ K to $10^{4.5}$ K. The SEDs in the bottom panel are all produced with a black hole mass of$\rm 10^{7.5} \msun$, and bolometric luminosities ranging from $\rm L_{bol} = 10^{42.5} erg/s$ to $10^{45.0} erg/s$, which correspond to characteristic temperatures of $10^{4.5}$ K to $10^{5.1}$ K. Additionally, in the bottom panel, we present the spectra normalized by their bolometric luminosity ($\rm \nu L_{\nu} / L_{bol}$) so that their shapes and features are easier to compare.}
    \label{fig:AGN_SED_const_M_lbol}
\end{figure}

\subsection{Mock Observation Pipeline}
\label{subsec:Obs_pipeline}

We use a nearly identical mock-observation process described in \citet{LaChance_2025} to calculate photometric magnitudes for each source. The one adjustment we make is to the AGN SED model, which is described in section \ref{subsec:AGN_SED_production}. We begin by calculating the intrinsic stellar emission of each star particle with the Binary Population and Spectral Population Synthesis model \citep[BPASS version 2.2.1;][]{Stanway2018}. These spectra are subject to dust attenuation via the interstellar medium (ISM), which we calculate using a simple power law dust model: 
\begin{equation} \label{eq1}
\tau_{\rm ISM}(\lambda) = -\kappa_{\rm ISM} \Sigma(x,y,z) \left(\frac{\lambda}{0.55 \mu m}\right)^{\gamma}
\end{equation}
where $\tau_{\rm ISM}(\lambda)$ is the optical depth, $\kappa_{\rm ISM}$ is an attenuation strength tuning parameter that we assign a value of $10^{4.1}$ \citep{astrid_galaxy_formation}, $\Sigma(x,y,z)$ is the metal surface density for each star, and $\gamma$ is the slope of the dust model. We set $\gamma = -1.0$, which falls between the slopes of the Small Magellanic Cloud model \citep{Pei_1992} and the `starburst' model \citep{Calzetti_2000}. We calculate the metal surface density for each star by creating a three dimensional map of the metal density. The map is smoothed with a 3D Gaussian kernel ($\rm \sigma = 1h^{-1} kpc$) to approximate the SPH kernel associated with each gas particle. The metal surface density of each star particle is the sum of the metal density in each of the voxels in front of it along the line of sight. For further details, see \citet{Wilkins17, Lachance_24}.

Young stars (Age $\leq 10$ Myr) are also subject to stellar birth cloud attenuation and can produce nebular emission. The nebular emission is produced following the procedure outlined in \citet{wilkins_2020}, and the birth cloud attenuation follows the process detailed in \citet{Vijayan_2021}. We use \texttt{SynthObs} \citep{wilkins_2020} to implement the entire stellar portion of the pipeline.

We calculate the observed magnitude in each band by summing the luminosities of all star particles and black holes within twice the stellar half-mass radius, thereby approximating the aperture used in observation while reducing compute time by skipping the full imaging process for each source. For the sources that meet the color criteria described in section \ref{subsec:LRD_identification}, we create a mock observed image to calculate the compactness of the source. We do so following the process described in \citet{Lachance_24} for the creation of mock CEERS observations. We produce these images at the drizzled resolution of 0.030'' (original long and short wavelength resolutions of 0.063'' and 0.031'' respectively), and apply a gaussian PSF to each filter, based on their empirical PSF widths reported for JWST (e.g. FWHM=0.145'' for the F444W filter, and 0.040'' for the F115W filter).

\subsubsection{AGN SED production}
\label{subsec:AGN_SED_production}

The AGN SED models in this work are based on the models from \citealt{LaChance_2025}~(with variations discussed in the next subsection), created with the \texttt{CLOUDY} \citep{cloudy} photoionization code. These models use the same AGN input SED with default parameters for the X-ray to UV ratio ($\alpha_{\rm ox}=-1.4$), the low energy slope of the Big Bump ($\alpha_{\rm UV}=-0.5$), and the slope of the X-ray component ($\alpha_X=-1.0$), and are run with characteristic temperatures ranging from $\rm 10^{4} K$ to $\rm 10^{6} K$, which are used to determine the AGN SED for a given black hole. The characteristic temperature of the AGN SED in \texttt{CLOUDY} corresponds to the peak of the ``blue bump'' in the incident SED. Assuming a standard optically thick, geometrically thin, accretion disc around a non-spinning black hole, the temperature at a given radius in the accretion disc is given by
\begin{equation} \label{eq2}
T_{\rm BH}(r) = \left(\frac{3 G M_{\rm BH}\dot{M}_{\rm BH}}{8\pi\sigma r^3}\right)^{1/4} (1-\sqrt{r_{ISCO}/r})^{1/4}
\end{equation}
where $T_{\rm BH}(r)$ is the temperature at distance r from the black hole, $M_{\rm BH}$ and $\dot{M}_{\rm BH}$ are the mass and accretion rate of the black hole respectively, $G$ is the gravitational constant, $\sigma$ is the Stefan–Boltzmann constant, and $r_{ISCO}$ is the radius of the inner most stable circular orbit for the black hole  ($r_{ISCO} = \frac{6 G M_{BH}}{c^2}$). This equation is maximized at a radius of $r = \frac{49}{36} r_{ISCO}$, and we use the temperature at this radius as the characteristic temperature for the AGN SED in \texttt{CLOUDY}.  
When evaluated at this radius ($\rm r = \frac{49}{36} r_{ISCO} = \frac{49 G M_{BH}}{6 c^2}$) the characteristic temperature scales with the mass and accretion rate of the black hole according to 
\begin{equation} \label{eq_T_BH_scaling}
T_{BH}(\frac{49}{36} r_{ISCO}) \propto M_{\rm BH}^{-1/2}, \dot{M}_{\rm BH}^{1/4}
\end{equation}

We set the metallicity of the gas cloud to 10\% the solar metallicity, and the inner cloud radius to $\rm log_{10}(r / cm)= 16.7$. All SEDs are produced at a set bolometric luminosity of $10^{43.5} \rm erg/s$. The SED of each black hole in the simulation is produced by calculating their characteristic temperature, creating an interpolated SED at that temperature, and then uniformly scaling that SED by the ratio between the black hole's luminosity and our reference of $10^{43.5} \rm erg/s$. We calculate the luminosity of each black hole with the equation 
\begin{equation} \label{eq_L_bol}
L_{\rm bol} = \eta\dot{M}_{BH} c^2
\end{equation}
where $\eta$ is radiative efficiency, for which we use the assigned values of $\eta=0.2~\&~0.1$ for \brahma\ and \astrid\ respectively.

We use equation \ref{eq1} to calculate the dust attenuation of the AGN due to the interstellar medium (ISM), but use a value of $\gamma = -2.0$ rather than $-1.0$. Dust attenuation plays a very minimal role in the emission of AGN in \brahma, as we shall show later, but we retain this AGN-specific adjustment to remain consistent with \citet{LaChance_2025}. The metal surface density for each black hole is calculated along the line-of-sight, using the SPH kernels of the gas particles in its host galaxy, following the procedure of \citet{astrid_BHs}. This process is more computationally expensive than using the metal density map we create for the stars, but produces slightly more accurate values, which are impactful given the relative brightness of AGN and individual star particles.

\subsubsection{Standard vs Gas-Enshrouded AGNs}

We explore two distinct AGN SED models which differ in their gas properties, but maintain a plane-parallel gas geometry. The first, hereafter referred to as ``standard AGN SED", is the exact AGN SED model used in \citet{LaChance_2025}. It has a hydrogen column density of $\rm log_{10}(N(H) /cm^{-2}) = 23.0$, a hydrogen volume density of $\rm log_{10}(n(H) / cm^{-3})= 9.0$, and no gas turbulence. While this is the more conservative of the two models we analyze, it does still include some assumptions that may be optimistic. Specifically, we implicitly assume a broad line region (BLR) cloud is present along the line-of-sight, \rev{and that this line-of-sight does not pass through the dusty torus. } %\sout{when some lines-of-sight, especially those along the polar regions, would not feature such dense gas clouds.} 
Additionally, our assumption of a metallicity of 10\% the solar metallicity, and minimal dust attenuation in the immediate AGN neighborhood is not guaranteed. As discussed in \citet{Inayoshi_2025}, there is a possibility of dust grains surviving in clouds within the dust sublimation radius. As such, there is a possibility that the \rev{number of standard AGN LRDs in \brahma, and \astrid\ could be even lower than we present here (We quantify these assumptions in Section~\ref{subsec:LRD_identification}).}

The second is a gas-enshrouded AGN SED model. It features an increased hydrogen column density of $\rm log_{10}(N(H) /cm^{-2}) = 24.5$, and hydrogen volume density of $\rm log_{10}(n(H) / cm^{-3})= 9.5$, similar to the models detailed in \citet{Inayoshi_2025}. This iteration of the gas-enshrouded AGN model is motivated by the increasing importance of the Balmer break to the spectral profile of LRDs \citep{Liu_2025}, and the similar gas-enshrouded AGN models used in \citet{Naidu_2025, Taylor_2025, Jeon_2025}. In addition to the increased gas density, and depth, which are implicit to the gas-enshrouded model, and necessary to produce the significant Balmer break in observed LRDs, turbulence~(with velocity $\rm v_{turb} = 400$ km/s) is added to the dense gas clouds in order to reproduce their observed features. Specifically, the width of the $\rm H\alpha$ and $\rm H\beta$ absorption lines, and the sloped Balmer breaks seen in observed LRD spectra \citep{Taylor_2025, Naidu_2025}. The exact values of these gas parameters vary between works, with $\rm n(H)=10^8 - 10^{11} cm^{-3}$, $\rm N(H) = 10^{23} - 10^{26} cm^{-2}$ and $\rm v_{turb} = 300 - 500$ km/s.

We decide on the above values for n(H) and N(H) based on a coarse parameter search of values for these two parameters. We present the results of this search in table \ref{table:AGN_param_grid}, with values of $\rm log_{10}(n(H) / cm^{-3})= 9.5$ and $\rm log_{10}(N(H) /cm^{-2}) = 24.5$ maximizing the number of LRDs in both \astrid\ and \brahma. Slight deviations from these values ($\pm 0.5$ dex) have fairly minimal impact on the number of LRDs found. More significant departures result in significantly fewer LRDs being identified.

\begin{table}
    \centering
    \begin{tabular}{|c||c|c|c|c|c|c|}
    \multicolumn{7}{c}{BRAHMA}\\
\hline
    & \multicolumn{6}{|c|}{$\rm log_{10}(N(H)/ cm^{-2}$)}\\
    \cline{2-7}
     $\rm log_{10}(n(H)/ cm^{-3}$) & 22.0 & 23.0 & 24.0 & \textbf{24.5} & 25.0 & 26.0\\
    \hline
    8.0 & 0 & 0 & 18 & 24 & 16 & 14 \\
    9.0 & 0 & 3 & 39 & 38 & 29 & 27 \\
    \textbf{9.5} & 1 & 8 & 40 & \textbf{41} & 37 & 30 \\
    10.0 & 1 & 8 & 38 & 40 & 39 & 34 \\
    11.0 & 1 & 4 & 27 & 33 & 36 & 36 \\
    \hline
    \multicolumn{7}{c}{ }\\
    \multicolumn{7}{c}{ASTRID}\\
    \hline
    & \multicolumn{6}{|c|}{$\rm log_{10}(N(H)/ cm^{-2}$)}\\
    \cline{2-7}
     $\rm log_{10}(n(H)/ cm^{-3}$) & 22.0 & 23.0 & 24.0 & \textbf{24.5} & 25.0 & 26.0\\
    \hline
    8.0 & 34 & 34 & 47 & 56 & 54 & 54 \\
    9.0 & 34 & 40 & 72 & 70 & 67 & 63 \\
    \textbf{9.5} & 37 & 43 & 74 & \textbf{74} & 70 & 68 \\
    10.0 & 37 & 44 & 70 & 73 & 70 & 69 \\
    11.0 & 38 & 42 & 59 & 62 & 62 & 62 \\
    \hline
    \end{tabular}
    \caption{The number of ``little red dots'' in the \brahma\ and \astrid\ simulations when applying AGN emission models with different values for the gas volume density, n(H), and gas surface density, N(H). The bolded values correspond to the gas-enshrouded model we used in this work ($\rm n(H) = 10^{9.5} cm^{-3}$, $\rm N(H)=10^{24.5} cm^{-2}$).}
    \label{table:AGN_param_grid}
\end{table}

A comparison between the standard AGN SED, and our iteration of the gas-enshrouded AGN SED is shown in Figure \ref{fig:example_spectrum} as the difference between the green and purple spectra. For standard AGN, the transmitted SED~(after being processed by the surrounding gas cloud) deviates minimally from the incident SED (black line) in the UV-visible range, featuring a very small Balmer Break and the production of emission lines. For the gas-enshrouded case, the transmitted AGN SED has a much more significant Balmer break, with wider absorption features due to the turbulence present in the gas. This large Balmer Break is sufficient to produce the red rest-frame optical color of an LRD as long as it is prominent compared to the AGN's host galaxy. 

In Figure \ref{fig:example_spectrum}, we also show multiple incident AGN SEDs at fixed luminosity but with different characteristic temperatures to illustrate the impact of temperature on the emitted spectrum (grey and black lines). The lower-temperature SED (dashed grey line, $\rm T=10^{4.5},K$) peaks at longer wavelengths than the higher-temperature SED (dotted grey line, $\rm T=10^{5.5},K$), making it brighter in the longer-wavelength JWST filters. Correspondingly, the slope of the AGN SED across the JWST-observable wavelength range shifts from red to blue with increasing temperature. In Figure \ref{fig:AGN_SED_const_M_lbol}, we show the transmitted AGN SEDs produced by the gas-enshrouded model for a range of black hole masses and bolometric luminosities. The trends observed in these spectra arise directly from the dependence of the characteristic temperature on black hole mass and luminosity, with lower temperatures corresponding to larger black hole masses and lower bolometric luminosities.

\begin{table}
    \centering
    \begin{tabular}{c|c|c}
    Simulation & standard AGN & gas-enshrouded AGN \\
    \hline
    \brahma\ & $9.9 \pm 7.0 \times 10^{-6}$    & $2.04 \pm 0.32 \times 10^{-4}$           \\
    \astrid\ & $8.4 \pm 2.0 \times 10^{-8}$    & $3.68 \pm 0.43 \times 10^{-7}$            \\
    \end{tabular}
    \caption{The number density of ``little red dots'' in the \brahma\ and \astrid\ simulations when applying the both the standard and gas-enshrouded AGN models described in section \ref{subsec:AGN_SED_production}. The models differ in their volume and surface density of Hydrogen, and gas turbulence. The ``standard'' model has a less dense, and less deep gas cloud surrounding the AGN than the gas-enshrouded model, and also lacks the turbulent motion present in the gas-enshrouded model.%The ``standard'' and ``gas-enshrouded'' models have values of $\rm log_{10}(n(H) / cm^{-3})= 9.0, \space 9.5$, $\rm log_{10}(N(H) /cm^{-2}) = 23.0, \space 24.5$, and $\rm \rm v_{turb} = 0, \space 400$ km/s respectively. 
    The standard model has gas properties of $\rm log_{10}(n(H) / cm^{-3})= 9.0$, $\rm log_{10}(N(H) /cm^{-2}) = 23.0$ and no turbulent velocity, while the gas-enshrouded model has $\rm log_{10}(n(H) / cm^{-3})= 9.5,~log_{10}(N(H) /cm^{-2}) = 24.5,~and~v_{turb} = 400~km/s$. All values in the table have units of $\rm [Mpc^{-3}]$, with Poisson uncertainties.}
    \label{table:LRD_num_dens}
\end{table}

\section{Results}
\label{sec:results}

We apply the mock-observation pipeline from Section \ref{subsec:Obs_pipeline} to generate synthetic images and spectra for simulated AGN–galaxy populations. We then identify sources that satisfy the criteria for classification as “little red dots” (LRDs), and analyze both the full AGN population and the LRD subset to determine which physical properties cause an AGN–host galaxy system to appear as an LRD.

\begin{figure*}
    \centering
    \includegraphics[width=2.0\columnwidth]{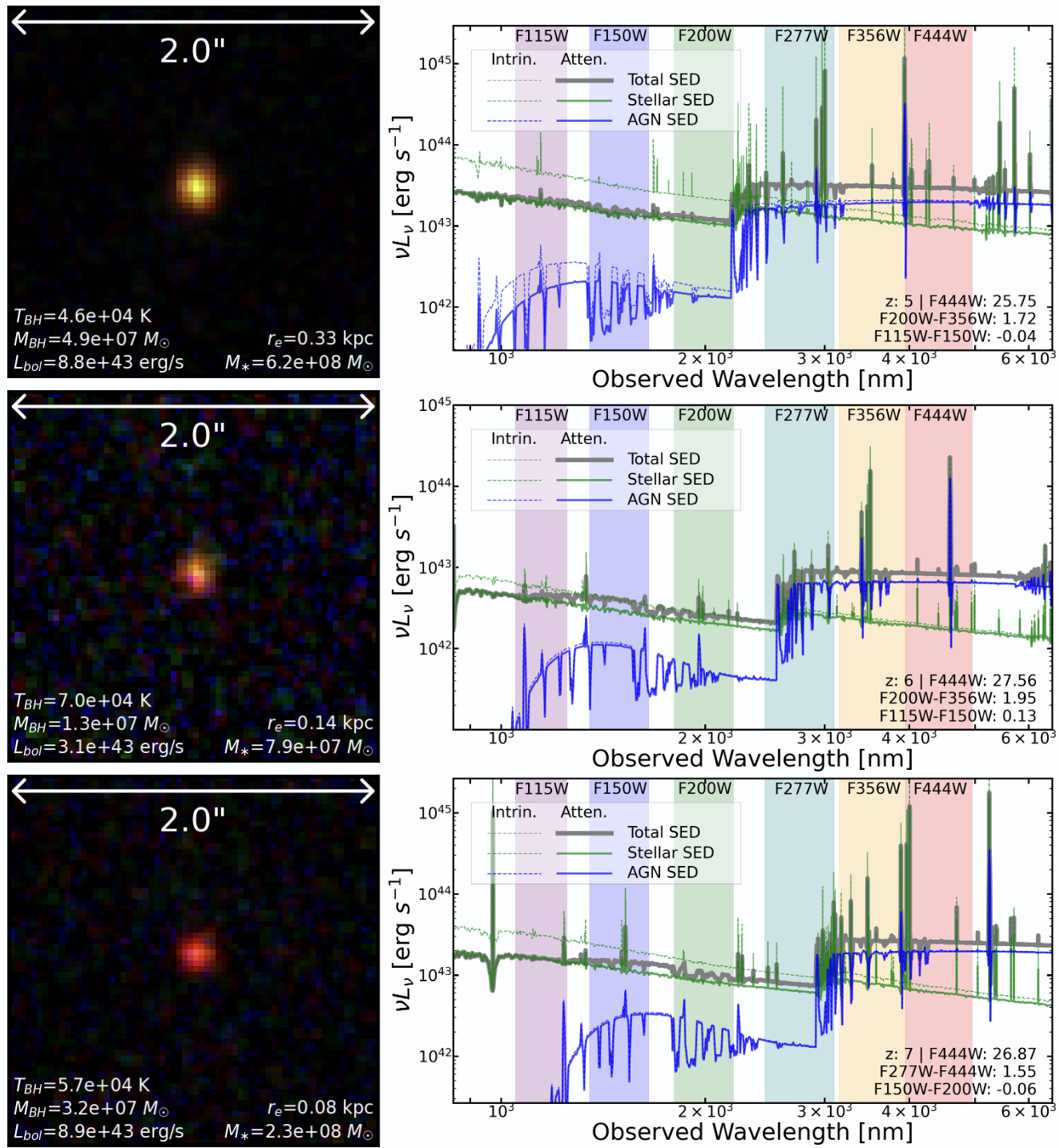}
    \caption{Mock JWST observations of three of the ``Little Red Dots'' found in the \brahma\ simulation. The images on the left are produced with the F444W, F277W, and F150W filters as the red, green, and blue channels, respectively. The spectra on the right show the total source spectrum in grey, the stellar component in green, and the AGN component in blue. The dashed lines indicate the spectra before dust attenuation is applied. On each mock image we include the properties of the black hole, and host galaxy. The black hole properties are the characteristic temperature for \texttt{CLOUDY}, the mass of the black hole, and its bolometric luminosity. The host galaxy properties are the Sérsic effective radius, and stellar mass.}
    \label{fig:Brahma_Mock_LRD_Obs}
\end{figure*}

\subsection{LRD Identification and Abundances} 
\label{subsec:LRD_identification}

We use LRD criteria similar to those used by UNCOVER \citep{Greene_2024, Labbe_2025}. We apply a magnitude cut on the F444W emission of $m_{F444W} < 28.0$ to ensure the objects are cleanly detected. Sources are then selected based on their colors, which must fit at least one of the Red1 or Red2 criteria,

\begin{align*}
\text{Red1 = (F115W-F150W < 0.8) and} \\
\text{(F200W -  F277W > 0.7) and} \\
\text{(F200W - F356W > 1.0),} \\
\text{Red2 = (F150W-F200W < 0.8) and} \\
\text{(F277W -  F356W > 0.7) and} \\
\text{(F277W - F444W > 1.0)}
\end{align*}

and their compactness in the F444W band, which must meet the following criterion:

$$\rm compact = f_{F444W}(0.4\text{"})/f_{F444W}(0.2\text{"}) \leq 1.7$$

where $\rm f_{F444W}(0.4\text{"})$ is the flux detected in the F444W band inside an aperture with a diameter of 0.4". We include both the Red1 and Red2 criteria to reduce the risk of selection effects arising from the shifting of rest-frame emission across the observation bands. This is particularly important with regard to the observed wavelength of the Balmer break, as it produces a significant change in the colors of these objects as its location shifts across the F277W filter from redshift $z=5$ to $8$ ($\lambda_{\rm obs}\sim2.2~\mu{\rm m}$ to $\sim3.3~\mu{\rm m}$). Broadly, this results in sources at redshifts 5 and 6 being selected by the Red1 criterion, and sources at redshifts 7 and 8 being selected by the Red2 criterion. All sources are evaluated using both selection criteria; however, for presentation purposes, we highlight the Red1 selection and corresponding colors for sources at redshifts 5 and 6, and the Red2 selection for sources at redshifts 7 and 8. Additionally, we assign each source a ``Rest-visible Color'' value based on the criterion associated with its redshift. For sources at redshifts $z = 5$ and $6$, ``Rest-visible Color'' is defined as $\rm F200W - F356W$, while for sources at redshifts $z = 7$ and $8$, ``Rest-visible Color'' is defined as $\rm F277W - F444W$.

Under the standard AGN SED model, only two objects across the four \brahma\ snapshots spanning $z=5{-}8$ satisfy all of the LRD selection criteria. Although the \brahma\ volume is relatively modest, this number remains substantially below expectations based on current JWST measurements. Given the observed LRD number density of $\sim10^{-4}~\rm Mpc^{-3}$ at these redshifts \citep{Pacucci_Loeb_2025, Kocevski_2025}, one would expect approximately $4$--$5$ LRD analogs per snapshot within our $[36~\rm Mpc]^3$ simulation volume. Instead, we identify only two objects across all four snapshots combined, corresponding to an effective number density of $\sim10^{-5}~\rm Mpc^{-3}$. While this abundance is more than two orders of magnitude higher than that predicted by \astrid\ ($8\times10^{-8}~\rm Mpc^{-3}$; \citealt{LaChance_2025}), owing to the substantially more massive black holes present in \brahma, it nevertheless remains approximately an order of magnitude below current observational estimates. %\sout{It is possible that one or both of these objects may not appear as LRDs, even under standard AGN conditions, depending on the line-of-sight (which may not pass through a BLR cloud), and any potential dust attenuation in the immediate region of the AGN.} 
\rev{This predicted LRD number density is an upper estimate, as we make a few favorable assumptions within the standard AGN model regarding the chemistry and geometry of the AGN environment.}

\rev{Regarding the chemistry, we compute our primary \texttt{CLOUDY} templates with a very low metallicity ($Z = 0.1 ~Z_{\odot}$) due to the early epoch at which these objects are found \citep{Trefoloni_2025}. As this is not reflective of low-redshift AGN, we run a grid of standard AGN \texttt{CLOUDY} models with $0.1 \lesssim Z/Z_{\odot} \lesssim 10$, and find that both standard AGN LRDs meet the LRD criteria up to $Z \approx 3 ~Z_{\odot}$, and one up to $Z \approx 10 ~Z_{\odot}$. This is consistent with our expectations, as the clouds being modeled lie within the dust sublimation radius, so we do not include dust grains in the model. As such, the heightened metallicity primarily alters the emission line properties, rather than the continuum. We also test the potential impact of unresolved nuclear dust, including dust that may survive within the sublimation radius \citep{Inayoshi_2025}, and find that any unresolved dust would need to contribute $A_V \gtrsim 0.75$ for either of the detected LRDs to fail to meet the LRD criteria. This indicates that our results are fairly resilient to changes to our assumptions regarding the chemistry of the standard AGN model.}

\rev{We find that the geometry assumptions have a significantly higher potential impact. As we cannot fully resolve the nuclear region, we rely on low-redshift measurements of the BLR and dusty torus covering factors, which indicate a covering factor of $\sim 0.1$--$0.5$ for the BLR \citep{Netzer_1990, Peterson_2006, Dunn_2007, Gaskell_2007, Baskin_2014, Baskin_Laor_2018}, of which we adopt $\sim0.3$--$0.5$, given the low Eddington ratios of the \brahma\ AGN \citep{Ferland_2020}, and indications of larger covering factors at high redshift \citep{Maiolino_2025}, and a covering factor of $\sim0.3$--$0.7$ for the dusty torus \citep{Mor_2009, RamosAlmeida_2011, Mor_Netzer_2012, Stalevski_2016, Ricci_2023}. In order for these objects to appear as LRDs the line-of-sight needs to both pass through a BLR cloud, and not intersect the dusty torus. This has a probability of $\sim 0.1$--$0.35$ indicating the expected number of standard-model LRDs may be $\sim3$--$10$ times lower than we estimate, if the covering factors of high-redshift AGN resemble those at low redshift.}

\begin{figure*}
    \centering
    \includegraphics[width=2.0\columnwidth]{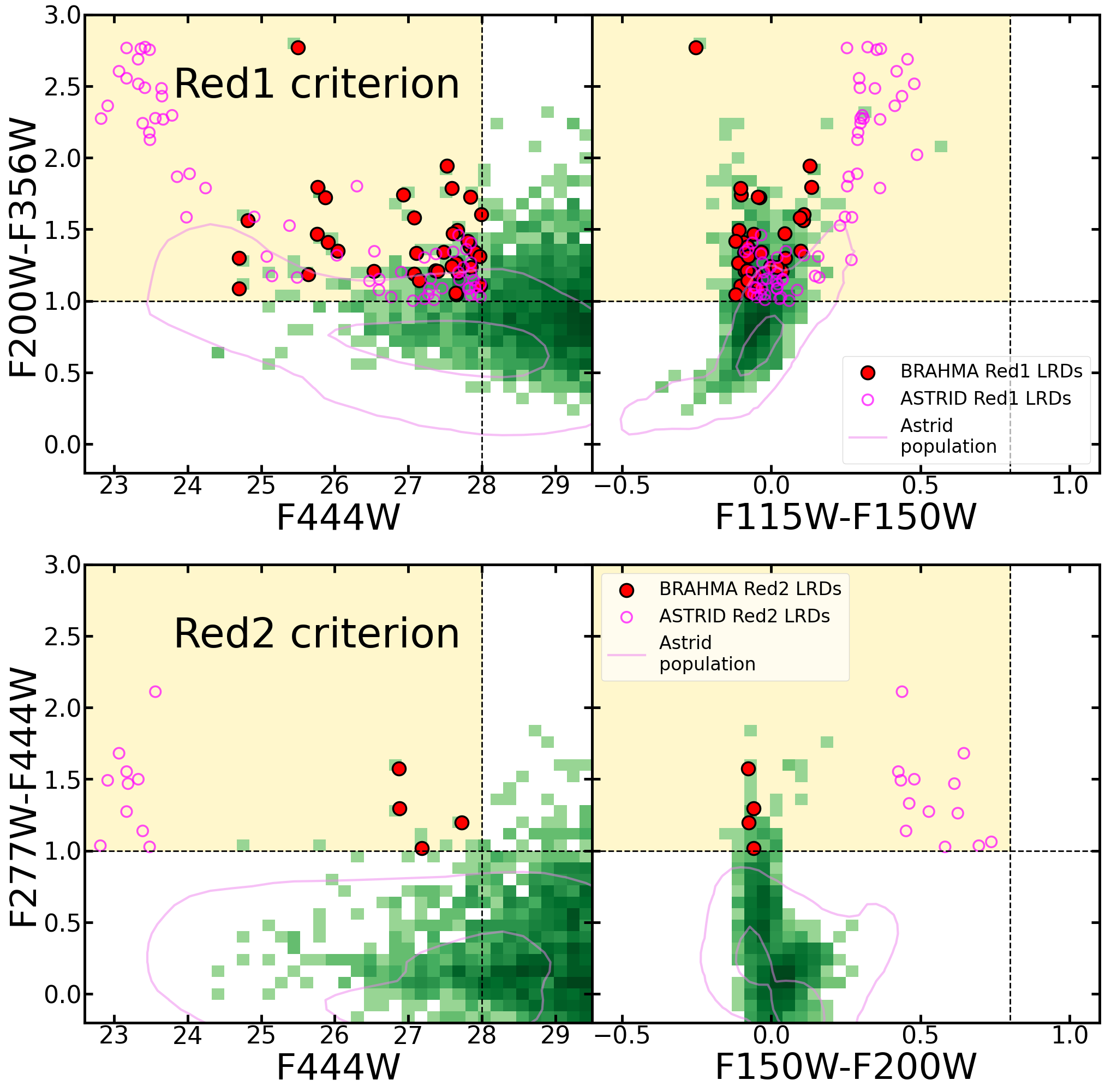}
    \caption{Color-Color and Color-Magnitude plots for the colors and magnitudes used in the ``little red dot'' criteria. The top panel shows sources F200W-F356W color vs. F444W magnitude, and F115W-F150W color. These are the primary factors in determining if a source meets the Red1 LRD criterion. The bottom panel shows sources F277W-F444W color vs. F444W magnitude, and F150W-F200W color. These are used in the Red2 LRD criterion. In order for a source to be considered a Red1 LRD it must be above the horizontal dashed line, and to the left of the vertical dashed line (falling within the highlighted regions) in both plots in the upper panel. The same is true for the Red2 criterion, and the plots in the bottom panel. The green 2D histogram shows the population of sources in \brahma. \contourtext. The LRDs identified in \brahma\ are marked with red circles, and the LRDs found in \astrid\ are shown with open magenta circles. Both panels include all of the sources in the snapshots we analyzed ($\rm z=5-8$).}
    \label{fig:LRD_criteria}
\end{figure*}

Next, we assign a gas-enshrouded AGN SED to every AGN in the \brahma\ volume and reapply the LRD selection criteria. We identify 41 LRD analogs across the four snapshots spanning $z\sim5{-}8$, representing a $\sim10$-fold increase relative to the standard AGN model. We show mock observations of three representative LRD analogs in Figure \ref{fig:Brahma_Mock_LRD_Obs}. Visually, all three sources appear compact and are dominated by a bright central emission component. Their spectra highlight the importance of the Balmer break, which is primarily responsible for the red colors observed in the long-wavelength JWST bands. In all three cases, the AGN dominates the emission redward of the Balmer break, while the host galaxy dominates the emission blueward of the break. These systems therefore fall into the hybrid LRD category, in which different regions of the rest-frame UV--optical spectrum are dominated by either the AGN or the host galaxy. In all three sources there is minimal dust attenuation, especially in the rest-frame visible range. The stellar components show slightly more dust attenuation than the AGN due to the attenuation of young stars from the remnants of their birth cloud. We present the color selection criteria applied to the entire population of \brahma\ sources in Figure \ref{fig:LRD_criteria}.

Amongst the \LRDnum\ gas-enshrouded LRD analogues in \brahma, \RedOneLRDnum\ of them fit the Red1 color criterion, and \RedTwoLRDnum\ fit the Red2 criterion.  We find an additional \notCompactLRDnum\ sources which meet the color criteria, but are too extended, and thus don't meet the compactness criterion. With the standard AGN model, we find two LRDs in \brahma\ both of which meet the Red1 criterion only. There are two additional objects which meet the Red1 criterion, but are too extended to be considered LRDs. 

However, recall that by assigning a gas-enshrouded SED to every AGN in the \brahma\ snapshots, we are effectively assuming an extreme, near-unity duty cycle for the gas-enshrouded phase. Consequently, the resulting LRD abundance of $\sim2\times10^{-4}~\rm Mpc^{-3}$ should be interpreted as an absolute upper limit on the number density of LRDs that can be produced by the \brahma\ AGN population. Even under these optimistic assumptions, this upper limit exceeds current observational estimates by only a factor of $\sim2$, implying that reproducing the observed LRD abundance would require gas-enshrouded duty cycles approaching unity. Such high duty cycles appear difficult to reconcile with current theoretical expectations~($\sim1\%$ in \citealt{Sun_2026}).

\begin{figure}
    \centering
    \includegraphics[width=1.0\columnwidth]{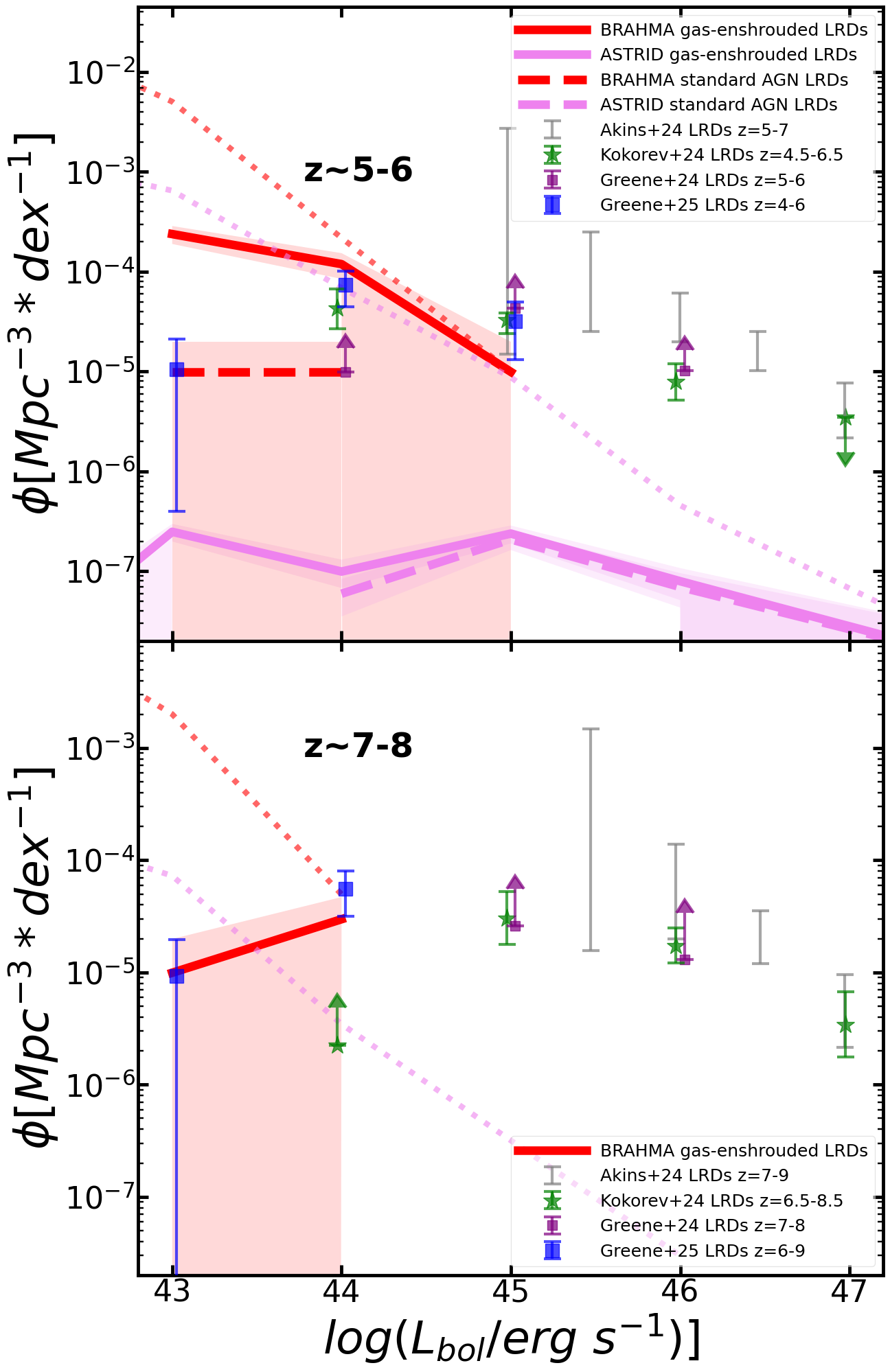}
    \caption{The bolometric luminosity function of the AGN for datasets around $5 \leq z \leq 6$ (top panel), and $7 \leq z \leq 8$(bottom panel). The \brahma\ and \astrid\ LRD AGN are shown in red and magenta respectively, with poisson uncertainty indicated with the shaded regions. The solid lines are the LRD populations with the gas-enshrouded AGN, and the dashed lines show the standard AGN LRD populations (Note, the standard AGN population in \brahma\ is only 2 objects). The overall black hole population for both simulations are shown with the corresponding dotted lines. We compare with the LRD AGN luminosity functions of multiple observations \citep{Kokorev_2024, Akins_2025} including both pre and post correction sets from \citet{Greene_2025}.}
    \label{fig:LRD_LF}
\end{figure}

We further compare the observed and simulated LRD abundances as a function of bolometric luminosity and redshift using the luminosity functions shown in Figure~\ref{fig:LRD_LF}. When conventional bolometric corrections are adopted \citep{Kokorev_2024, Greene_2024}, the \brahma\ LRD analogs are systematically less luminous than the observed population. However, \citet{Greene_2025} recently proposed an alternative bolometric correction for high-redshift AGN that shifts the inferred luminosity function by approximately one dex toward lower luminosities. Under this new bolometric correction, the \brahma\ upper limits are in good agreement with the observed estimates at $z\sim7{-}8$ for $L_{\rm bol}\sim10^{43}-10^{44}~\rm erg~s^{-1}$ and at $z\sim5{-}6$ for $L_{\rm bol}\sim10^{44}-10^{45}~\rm erg~s^{-1}$. Only at $z\sim5{-}6$ and $L_{\rm bol}\sim10^{43}~\rm erg~s^{-1}$ do the \brahma\ upper limits exceed the \citet{Greene_2025} measurements by a factor of $\sim20$. However, these luminosities correspond to the faintest sources currently accessible to JWST, where observational incompleteness may represent a significant source of uncertainty.

For the \astrid\ simulation, adopting the gas-enshrouded AGN model yields an upper limit on the LRD abundance of only $3.7\times10^{-7}~\rm Mpc^{-3}$, approximately $2$--$3$ orders of magnitude below current observational estimates. This again reflects the substantially lower black hole masses in \astrid\ relative to \brahma, reinforcing the conclusion that overmassive black holes may be a necessary, though not sufficient, for producing the observed abundance of LRDs.

\begin{figure}
    \centering
    \includegraphics[width=1.0\columnwidth]{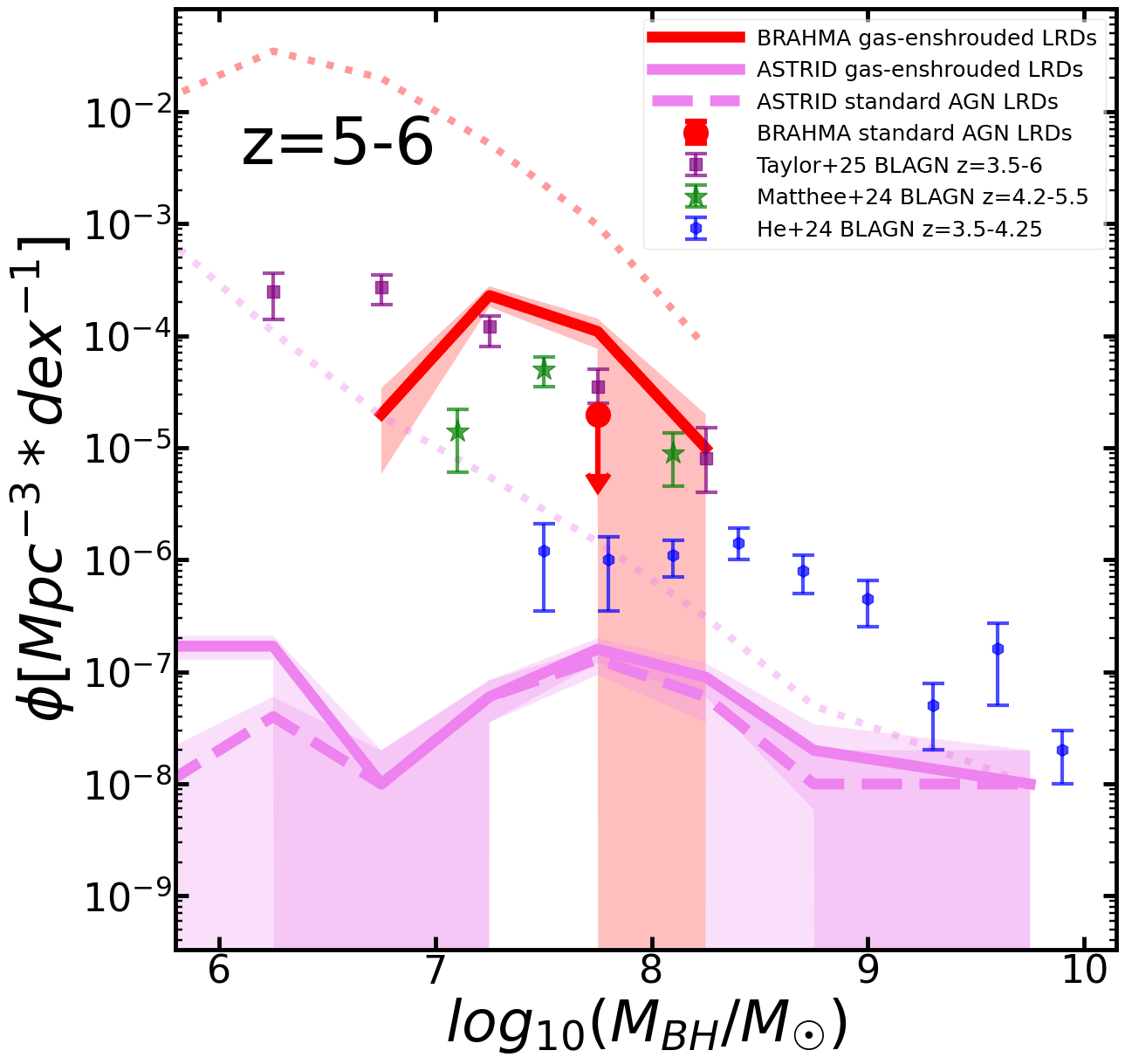}
    \caption{The black hole mass function of LRDs in \brahma\ and \astrid\ compared with observed black hole mass functions for all broad-line AGN (BLAGN) up to  $z=6$ \citep{Taylor_2025b, Matthee_2024, He_2024}. The AGN found in LRDs in \brahma\ are shown in red, and the \astrid\ LRDs are shown in magenta. The solid lines are for the LRDs produced with the gas-enshrouded AGN SED, and the dashed for the standard AGN SED. We include a red point for the LRDs in the \brahma\ simulation with the standard AGN SED, as they fall within a single mass bin. We also include the overall black hole mass function of \brahma, and \astrid, as dotted lines. }
    \label{fig:BHMF_and_z_trend}
\end{figure}

In Figure~\ref{fig:BHMF_and_z_trend}, we also show the upper limits of the LRD black hole mass function predicted by \brahma\ using the gas-enshrouded AGN SED models. In this limit, the black hole masses of the \brahma\ LRDs span $\sim7\times10^{6}$--$2\times10^{8}~M_{\odot}$. A direct comparison with the observed broad-line AGN (BLAGN) mass functions~\citep{Taylor_2025b, Matthee_2024, He_2024} is not straightforward, as not all BLAGNs are classified as LRDs. However, in the \cite{Taylor_2025b} sample, approximately $30\%$ of the BLAGNs are identified as LRDs. Assuming a similar fraction applies more generally, the \brahma\ upper limits appear somewhat higher than the observational estimates at the massive end, although the statistics remain limited due to the small simulation volume. In contrast, the \brahma\ LRD mass function drops sharply below black hole masses of $\sim10^{7}~M_{\odot}$, whereas the observational constraints from \cite{Taylor_2025b} extend down to $\sim10^{6}~M_{\odot}$. This discrepancy could be alleviated if the fraction of BLAGNs classified as LRDs decreases toward lower black hole masses. If this is not the case, however, the tension between the observed LRD abundances and the \brahma\ predictions would arise primarily from the low-mass end, corresponding to black holes with masses $\lesssim10^{7}~M_{\odot}$. We caution, however, that this interpretation assumes the current observational black hole mass estimates can be taken at face value. Several studies have suggested that these masses may be systematically overestimated.

Overall, we find that the overmassive black holes present in \brahma\ produce substantially higher LRD abundances than those found in \astrid, where such a population is absent. However, reproducing the observed LRD abundance with the \brahma\ AGN population requires gas-enshrouded duty cycles approaching unity, which appear inconsistent with current theoretical expectations. This tension may be driven by the scarcity of low-mass ($\lesssim10^{7}~M_{\odot}$) LRDs in \brahma\ relative to what is implied by current JWST-based estimates of the black hole mass function for BLAGNs. However, both the LRD fraction among BLAGNs and the black hole mass estimates themselves remain uncertain. Nevertheless, we conclude that, while overmassive black holes can significantly enhance the abundance of LRDs, they are unlikely to fully account for the observed JWST LRD population on their own.

\subsection{Black hole-galaxy relations: LRD vs full population}
\label{subsec:BH_gal_props}

While the overmassive AGN population in \brahma\ is unlikely to fully account for the observed LRD abundances, it is nevertheless instructive to examine the LRD analogues identified in the simulation and compare them to the broader galaxy population. Here, we focus on the locations of the \brahma\ LRD analogues in the $M_*$--$M_{\rm BH}$, $M_*$--$L_{\rm bol}$, and $M_{\rm BH}$--$L_{\rm bol}$ planes, and compare them to the overall galaxy population in \brahma.

\begin{figure}
    \centering
    \includegraphics[width=1.0\columnwidth]{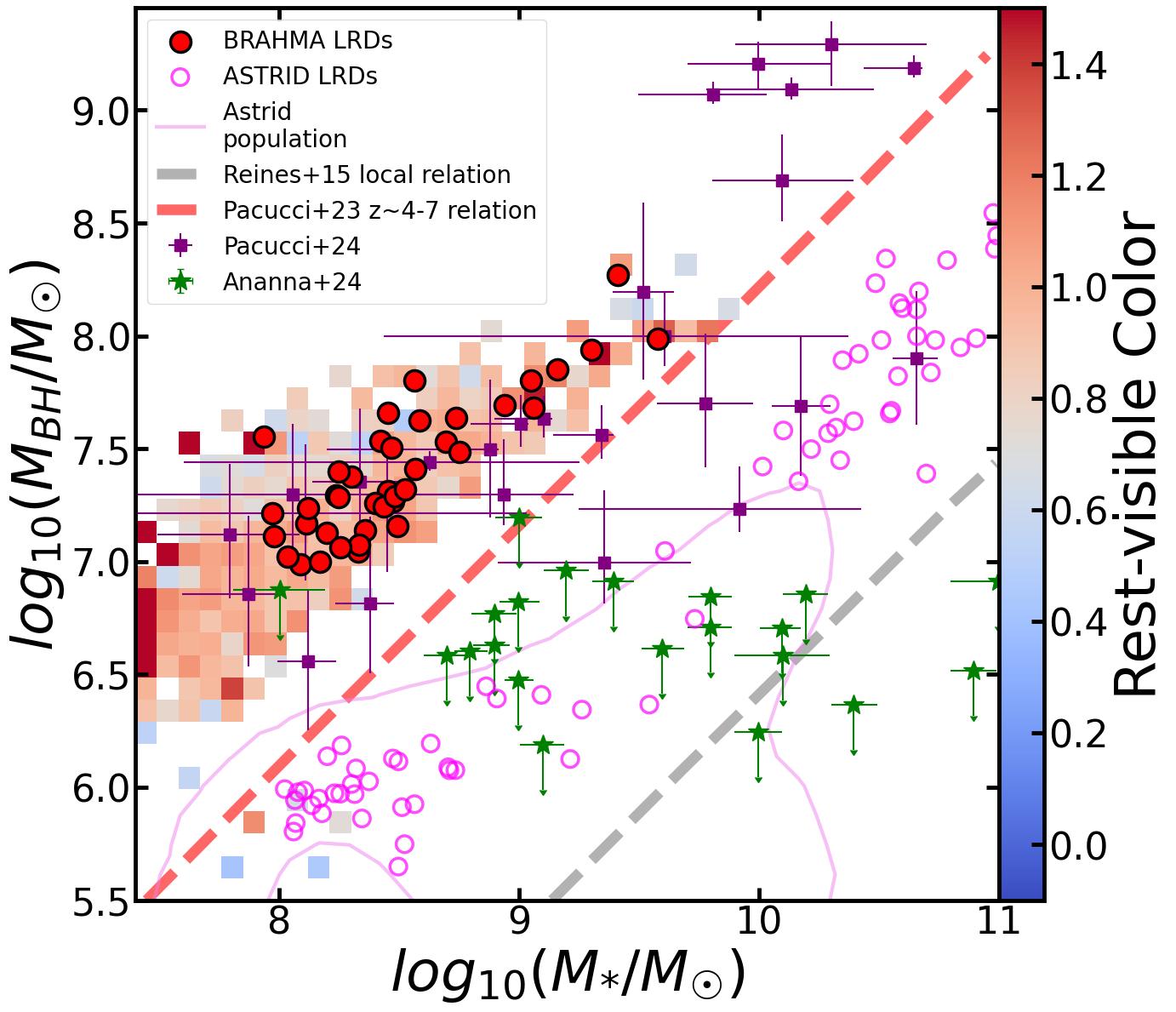}
    \caption{The $\rm M_{BH}-M_{\ast}$ for the black holes and their host galaxies in the \brahma\ and \astrid\ simulations. \RedColorText. \contourtext. The LRDs identified in \brahma\ are marked with red circles, and the LRDs found in \astrid\ are shown with open magenta circles. The purple square dataset is comprised of AGN fits from \citet{Maiolino_2024_JADES, Harikane_2023, Ubler_2023, Stone_2024, Furtak_2024, Yue_2024_eiger} compiled in \citet{Pacucci_Loeb_2024}. The green star dataset was produced in \citet{Ananna_2024} from the X-ray observations of some LRDs. We also include the local relation per \citet{Reines_2015} and an inferred $\rm z\sim 4-7$ relation from \citet{Pacucci_2023}}
    \label{fig:MBH_Mstar}
\end{figure}

In Figure~\ref{fig:MBH_Mstar} we show the populations on the $M_{\rm BH}$--$M_{\ast}$ plane. For galaxies of a given stellar mass, the \brahma\ population, including the LRDs, hosts black holes that overlap with the overmassive black holes inferred from \textit{JWST} observations~\citep{Pacucci_2023}. As discussed earlier, this agreement with the current observational constraints is the primary motivation for focusing on this particular \brahma\ realization. In contrast, black holes in the \astrid\ simulation are typically one to two orders of magnitude less massive than their \brahma\ counterparts at fixed stellar mass.

Within the \brahma\ population, there is minimal correlation between black hole mass and rest-visible color. Unlike \astrid, where the LRDs preferentially occupy the upper envelope of the $\rm M_{BH}$--$M_{\ast}$ relation~(Figure 4 of \citealt{LaChance_2025}), the LRDs in \brahma\ span a broad range of $\rm M_{BH}/M_{\ast}$ values and are not preferentially associated with the most overmassive black holes in the simulation. As already demonstrated, the presence of an overmassive black hole can have a significant impact on the likelihood of a galaxy appearing as an LRD, making the lack of correlation between black hole mass and redness in \brahma\ particularly noteworthy. We believe this arises because essentially all black holes in this \brahma\ simulation are already substantially more massive than the black holes hosted by similar-mass galaxies in \astrid. As a result, even some of the galaxies occupying the lower envelope of the \brahma\ $\rm M_{BH}/M_{\ast}$ distribution host black holes that are sufficiently massive for their AGN emission to make them appear as LRDs. Consequently, once a population of systematically overmassive black holes is established, variations in $\rm M_{BH}/M_{\ast}$ within that population become a secondary factor in determining whether a galaxy appears as an LRD.

\begin{figure*}
    \centering
    \includegraphics[width=2.0\columnwidth]{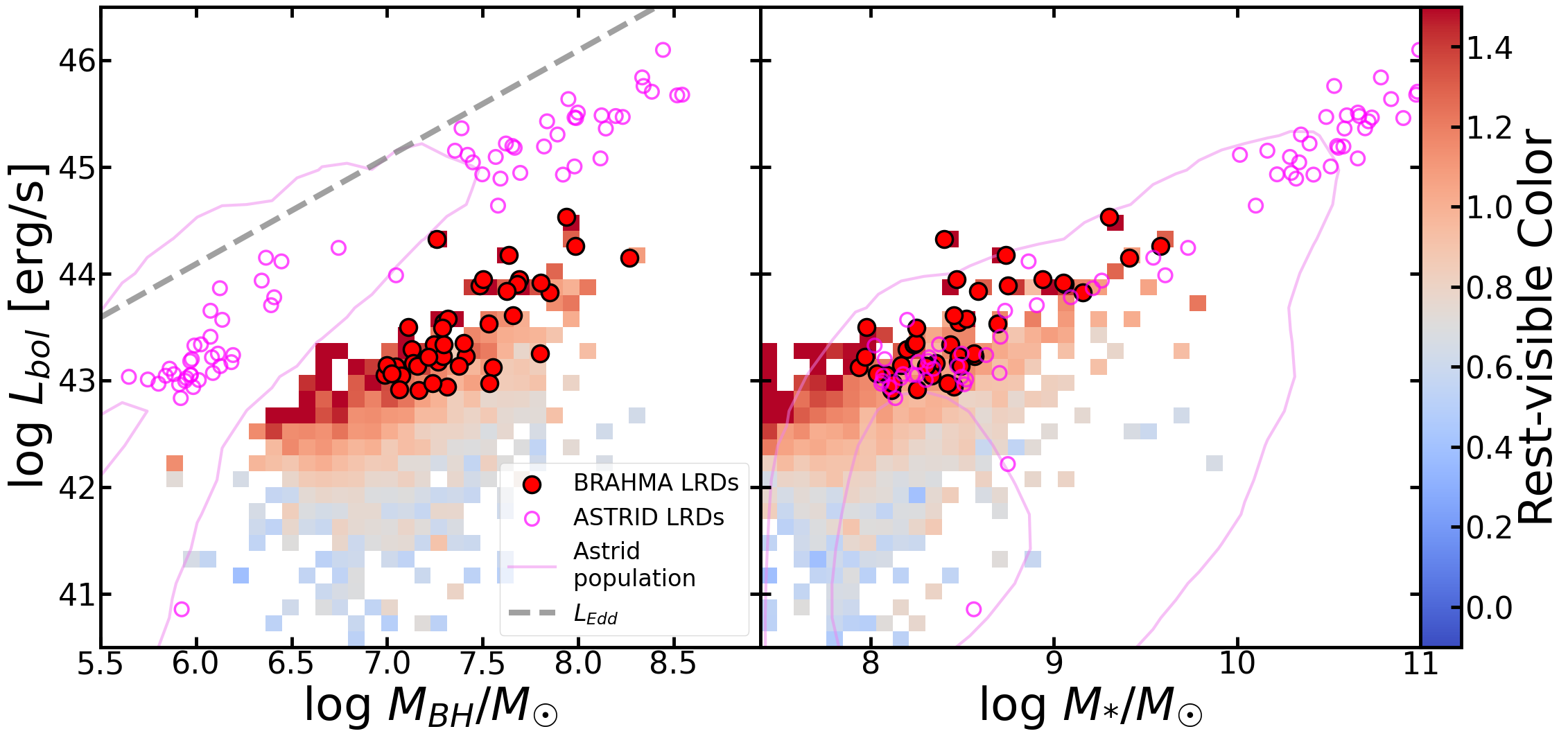}
    \caption{$\rm L_{bol}-M_{BH}$ and $\rm L_{bol}-M_{\ast}$ for the black holes and their host galaxies in the \brahma\ and \astrid\ simulations. \RedColorText. \contourtext. The dashed grey line in the left panel indicates the Eddington luminosity as a function of black hole mass.}
    \label{fig:Lbol_MBH_Mstar}
\end{figure*}

In Figure~\ref{fig:Lbol_MBH_Mstar} we show the bolometric luminosities of the black holes in our sample as a function of both black hole mass and host-galaxy stellar mass. The \brahma\ black holes are systematically dimmer at fixed black hole mass than their counterparts in \astrid, corresponding to significantly lower Eddington ratios. As mentioned earlier, these predominantly sub-Eddington accretion rates differ from the original application of the gas-enshrouded AGN model to LRDs by \citet{Inayoshi_2025}, which was associated with %motivated by
episodes of near- or super-Eddington accretion~(see also \citealt{ Inayoshi_2025b, Liu_2025, Naidu_2025, Taylor_2025}). If such accretion rates are indeed an inevitable consequence of the gas-enshrouded phase, then our application of the gas-enshrouded SED to the \brahma\ population would only be physically self-consistent if the associated LRD duty cycle is significantly smaller than unity. In this scenario, the brief gas-enshrouded phase could coincide with unresolved accretion bursts, while the time-averaged accretion rates tracked by the simulation remain sub-Eddington. But here again, recall that the observed LRD abundances can only be reproduced if we effectively assume near-unity duty cycles for the gas-enshrouded phase. This further highlights that the overmassive \brahma\ population is likely not sufficient to fully account for the observed LRD populations.

The brightness of black holes in galaxies of a given mass is relatively consistent between \brahma\ and \astrid\ due to the shift in the $\rm M_{BH}-M{\ast}$ relation shown in figure \ref{fig:MBH_Mstar}. There is a strong correlation between higher bolometric luminosity and ``Rest-visible Color'', with sources hosting brighter black holes being significantly redder, and all of the LRDs in \brahma\ containing black holes with a bolometric luminosity of $\rm log_{10}(L_{bol} / erg \space s^{-1}) \geq 42.9$. This is expected as the large Balmer break present in the gas-enshrouded AGN SED model will have a significant impact on the color for sources whose AGN contributes a larger portion of the visible light. 

Notably, there is significant overlap between the region of $\rm L_{bol}-M_{\ast}$ space where the LRDs found in \brahma\ lie, and the general \astrid\ population, and there are some LRDs found in \astrid\ there. While the total number of LRDs in this region appears comparable between the two simulations, \astrid\ has $1000\times$ more simulation volume than the \brahma\ simulation we analyzed, so the number density of LRDs in this region is $\sim 1000\times$ higher in \brahma\ than \astrid. 

The primary differentiating property of the \brahma\ sources is their significantly higher black hole mass relative to their \astrid\ counterparts, which results in AGN spectra with lower characteristic temperatures. %For a given luminosity, the higher mass black holes in \brahma\ will have AGN spectra with lower characteristic temperatures than the lower mass black holes in \astrid. 
For black holes with bolometric luminosities near those of the \brahma\ LRDs ($\rm 42.5\leq log_{10}(L_{bol}/(erg/s)) \leq 44.5$) the typical characteristic temperatures of \brahma\ and \astrid\ AGN are $\rm \sim10^{4.6} K$, and $\rm \sim10^{5.4} K$ respectively. This results in the \brahma\ AGN being nearly twice as bright in the F444W band as their \astrid\ counterparts.
This reinforces the importance of the overmassive black holes present in this \brahma\ simulation to the production of these dimmer LRDs, albeit not sufficient to explain the observed LRD abundances.


%\subsection{What Makes an LRD \aklant{in \brahma~ and \astrid}?}
\subsection{The role of AGN, host-galaxy emission, and dust attenuation in \brahma\ LRDs}
\label{subsec:LRD_properties}

\begin{figure*}
    \centering
    \includegraphics[width=2.0\columnwidth]{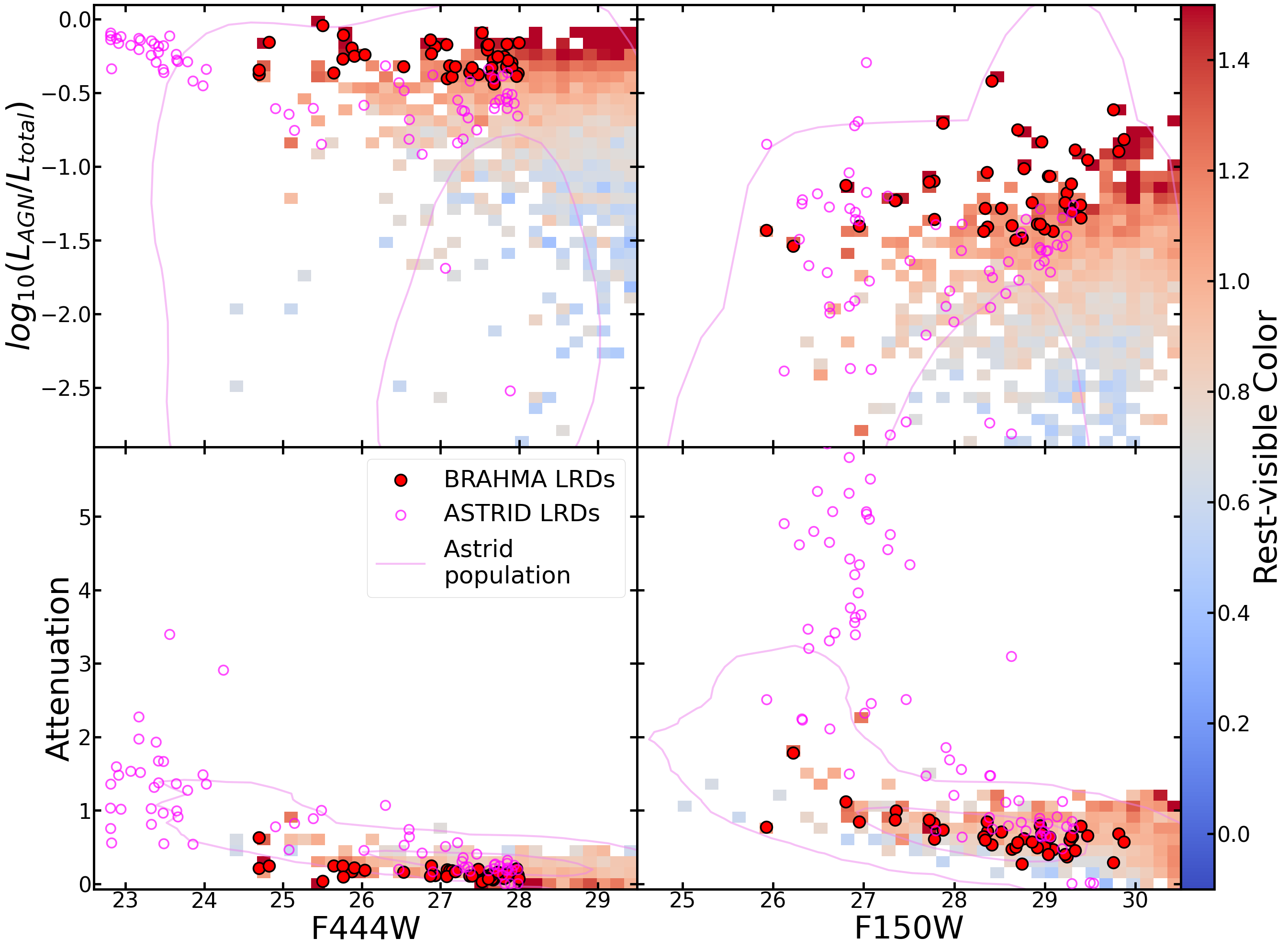}
    \caption{The top row shows the relative contribution of the AGN to the overall emission in both the F444W and F150W bands. The bottom row shows the amount of dust attenuation experienced by each source in the F444W and F150W bands. The x-axis of each plot is the source magnitude in the relevant filter (F444W for the left column, F150W for the right column). \RedColorText. \contourtext.}
    \label{fig:AGN_contribution_Attenuation}
\end{figure*}

Having characterized the black hole and host-galaxy properties of the LRD analogs identified in \brahma, we now investigate how emission from the AGN and host galaxy combine to produce their observed photometric and spectroscopic properties. In Figure~\ref{fig:AGN_contribution_Attenuation}, we quantify the relative contributions of the AGN and host galaxy, as well as the degree of dust attenuation, in two JWST bands that bracket the Balmer break: F150W and F444W. The LRD analogs in \brahma\ (as well as those in \astrid) generally exhibit significant AGN contributions in the F444W band, which probes rest-frame wavelengths redward of the Balmer break, while their F150W emission is typically dominated by the host galaxy. We previously highlighted this behavior in the example LRD spectra shown in Figure~\ref{fig:Brahma_Mock_LRD_Obs}; here we demonstrate that it is a generic feature of the full simulated LRD population.

In the lower panels of figure \ref{fig:AGN_contribution_Attenuation} we see that the \brahma\ LRD population generally shows very minimal attenuation, especially in the longer wavelengths, with mean and standard deviation of the attenuation in the F444W $\rm A_{F444W} = 0.15 \pm 0.09$. We also compute the rest frame visible attenuation based on the NIRCam filter centered nearest to the V-band (F356W for z=5,6 and F444W for z=7,8), and find $\rm A_{V} = 0.21 \pm 0.12$. 

This differs from the brighter \astrid\ LRDs, many of which feature $\rm A_{F444W} \geq 1.0$ (and $\rm A_{F150W} \geq 3.0$). This is indicative of the different reddening mechanisms that are contributing to the red color of the LRDs. These brighter LRDs host more massive galaxies, which can sufficiently attenuate their AGN to produce the red color associated with LRDs without a significant Balmer break. In contrast, the minimal dust attenuation present in the \brahma\ LRDs indicates they must be reddened in another way - in this case via the strong Balmer break of their gas-enshrouded AGN.

While we do not model dust re-emission in the current pipeline, it is extremely unlikely that the minimal amount of attenuated light in the \brahma\ LRDs would be sufficient to produce enough re-emission in the infrared to exceed the current observational limits. As a lower bound on the emission in this region we calculate the luminosity of one of the brightest AGN among the LRDs ($\rm log_{10}(M_{BH} / \msun)=7.94$, $\rm log_{10}(L_{bol} / erg\space s^{-1}) = 44.5$, z=6) in the ALMA 1.2mm band to be $\rm \nu L_{\nu}=5.9*10^{6} \lsun$, which is less than $1\%$ of the ALMA upper limits ($ \rm \nu L_{\nu} \sim 10^{9} \lsun$) \citep{Casey_2025, Setton_2025, Xiao_2025, Akins_2025} 

\section{Discussion}
\label{sec:discussion}

We have demonstrated that the black hole population in \brahma\ is unable to fully account for the observed abundance of LRDs. Interpreting this discrepancy requires understanding both the limitations of the underlying simulation model and the possibility that LRDs constitute a heterogeneous population with multiple distinct physical origins. We therefore review the scenarios proposed for the nature of LRDs and discuss which of them can be meaningfully investigated with the current \brahma\ simulations and which cannot.

\subsection{LRD scenario that is probed in this work}

We begin by describing the LRD scenario that is directly probed by our analysis. To do so, we revisit the masses and growth histories of the LRD analogs identified in \brahma, as well as the assumptions underlying our AGN modeling. First, the black holes hosting our LRD analogs typically have masses exceeding $\sim10^{7}~M_{\odot}$, more than two orders of magnitude larger than the characteristic seed mass of $\sim10^{5}~M_{\odot}$ adopted in this realization. Second, as discussed previously, the growth of these black holes is dominated by the initial seed mass and subsequent black hole mergers, with gas accretion playing a comparatively minor role. This behavior is a natural consequence of selecting a \brahma\ realization that produces abundant heavy seeds and efficiently merges them through the repositioning scheme. Consequently, the LRD analogs identified in our simulations do not probe the formation of the first black hole seeds or their earliest growth phases, but instead correspond to transient gas-enshrouded phases within a much later evolutionary stage in which the black holes have already undergone substantial merger-driven growth. The resulting overmassive black holes have lower temperatures, which in turn produce redder AGN spectra and increase the likelihood that the source satisfies the LRD color-selection criteria, particularly when coupled with the gas-enshrouded AGN model adopted in this work.

With that being said, reproducing the observed LRD abundance within this framework requires gas-enshrouded duty cycles approaching unity, which appear difficult to reconcile with current theoretical expectations \citep{Sun_2026}. This motivates consideration of additional LRD formation scenarios that are not captured by our present modeling and may contribute significantly to the observed population.

\subsection{Other LRD scenarios and modeling limitations in BRAHMA}

Several LRD formation scenarios have been proposed in the literature that are difficult to investigate within the current \brahma\ simulations due to resolution and modeling  limitations. We briefly discuss these possibilities below and assess how they may affect the interpretation of our results.

At the time of their initial discovery, two broad classes of explanations were considered: ultracompact star-forming galaxies \citep{Baggen_2023, Barro_2023, Perez_Gonzalez_2024} and AGN-powered systems \citep{Labbe_2023, Labbe_2024, Maiolino_2024_JADES, deGraaff_2025_BHstar}. The former cannot be robustly investigated within \brahma. In addition to the limited spatial resolution of the simulation, the interstellar medium is modeled using an effective equation of state, preventing the explicit treatment of the dense, multiphase gas structures and compact nuclear starbursts invoked in many stellar-powered LRD scenarios. While purely stellar explanations have become increasingly disfavored for at least a subset of the LRD population, due to the high stellar densities required \citep{Guia_2024}, and the low upper limits on dust \citep{Casey_2024}, %owing to the extreme stellar densities required and the difficulty of naturally reproducing the broad Balmer emission lines observed in some sources, 
they have not been conclusively ruled out. Consequently, if a significant fraction of the observed LRD population originates from ultracompact star-forming galaxies, those sources would not be represented among the LRD analogs identified in this work.

AGN-powered interpretations of LRDs have gained increasing support in recent years and were the primary motivation for this work \citep{Inayoshi_2025, Naidu_2025, Taylor_2025, Jeon_2025, deGraaff_2025_BHstar} However, even within the AGN framework, several distinct physical scenarios have been proposed, not all of which can be effectively investigated with simulations such as \brahma.

The first possibility within the AGN powered scenario is that LRDs are directly catching the light emitted during the formation the first black hole seeds, particularly direct collapse black holes \citep{Naidu_2025, Pacucci_2026_DCBH} Cosmological simulations like \brahma\ come no where close to actually resolving the seed formation process(es), while modeling the emission from them using post-processing is also difficult due to the small scale uncertainties in the environments associated with different seed formation scenarios. Therefore, if a sizable fraction of LRDs are indeed probing the formation of the first black hole seeds, our work does not include their contribution at all.   

Another closely related possibility is that LRDs are not directly probing black hole seed formation itself, but rather an immediate post-seeding phase of rapid~(hyper-Eddington) growth \citep{Pacucci_Narayan_2024, Inayoshi_2024, Lambrides_2024}. Such episodes may occur during the earliest growth stages of DCBH seeds and, in some cases, Pop III remnants. However, we do not observe such rapid accretion phases in the \brahma\ simulations, including the specific realization analyzed in this work. Instead, black hole growth in \brahma\ is typically highly sub-Eddington at early times, with Eddington ratios decreasing for lower-mass black holes. This may explain why we identify very few LRD analogs hosting black holes below $\sim10^{7}~M_{\odot}$ even with the gas enshrouded AGN model. This suppression of rapid early accretion is partly due to regulation from stellar and AGN feedback. At the same time, the effective equation-of-state treatment of the interstellar medium inhibits the formation of the extremely dense gas concentrations required for hyper-Eddington growth. Consequently, if a significant fraction of the observed LRD population is associated with rapid-growth phases immediately following seed formation, our analysis would entirely miss their contribution.

Overall, given the diversity of proposed LRD formation scenarios and the limitations of the physical models adopted in \brahma, it is not surprising that the specific LRD channel probed in this work—namely, transient gas-enshrouded accretion phases associated with overmassive ($\gtrsim10^{7}~M_{\odot}$) black holes—may not account for the entirety of the observed LRD population. Nevertheless, it is encouraging that this scenario is capable of reproducing at least a subset of the observed LRDs, provided that the gas-enshrouded phase occurs with a sufficiently high duty cycle. Future work will explore the alternative LRD formation pathways discussed above, which will likely require simulations with substantially higher spatial resolution and a more detailed treatment of the multiphase interstellar medium and black hole environments.

\subsection{Comparison with other theoretical works}
The assembly of LRDs has also been investigated in other theoretical frameworks, including cosmological simulations~\citep{Cenci_2025} and semi-analytic models (SAMs)~\citep{Jeon_2025}. Here we place our results in the context of these studies, while noting that direct comparisons are complicated by differences in both the underlying physical models and the specific LRD formation scenarios considered.

A key distinction between our analysis and these previous works is the LRD selection methodology. We identify LRD analogs using the observed photometric color-selection criteria, whereas the A-SLOTH SAMs of \citet{Jeon_2025} define LRDs based on black hole properties. In their Pop III seed model, the LRD population consists of black holes with $M_{\rm BH}\gtrsim10^6~M_{\odot}$ accreting above the Eddington limit, while in their DCBH model it includes all black holes with $M_{\rm BH}\gtrsim10^6~M_{\odot}$. Under these assumptions, they find that Pop III seed models overpredict the observed LRD abundances, whereas DCBH seed models are broadly consistent with them.

Applying an analogous mass-based selection to our heavy-seed model would correspond to the dotted red curve in Figure~\ref{fig:BHMF_and_z_trend}, yielding black hole mass functions substantially above both the observed LRD abundances and the DCBH predictions of \citet{Jeon_2025}. This difference is not unexpected, however, because the heavy-seed model adopted in this \brahma\ realization is much more lenient, as it does not impose the additional Lyman--Werner flux based criteria employed by the DCBH model in \citet{Jeon_2025}. Once the observational color-selection criteria are applied, our predicted LRD black hole mass functions (upper limits) are significantly suppressed (solid versus dotted red curves in Figure~\ref{fig:BHMF_and_z_trend}), bringing them much closer to the observations for black holes with $M_{\rm BH}\gtrsim10^7~M_{\odot}$. Whether a similar suppression would occur in the models of \citet{Jeon_2025} under our adopted color-selection criteria remains an open question, as it would also depend on the black hole accretion rates and host-galaxy properties, both of which may differ substantially between their SAMs and our cosmological simulations.

\citet{Cenci_2025} use the MELIORA simulation to investigate an LRD formation pathway that features emission from a ``young DCBH''~\citep{Inayoshi_2025b} and its host galaxy. (see also \citealt{Pacucci_2026_DCBH} for a ``young-DCBH''-only model). %\citep{Naidu_2025, Pacucci_2026_DCBH}
They find that a substantial fraction of newly formed DCBH seeds undergo an AGN phase reaching luminosities of $\gtrsim10^{43}~\rm erg~s^{-1}$ within $\lesssim200~\rm Myr$ of their formation. In contrast, black holes in our \brahma\ realization do not attain such luminosities on these timescales. Instead, they reach comparable luminosities only after growing to masses of $\gtrsim10^7~M_{\odot}$, primarily through mergers. Therefore, as discussed in the previous subsection, our \brahma\ simulation does not produce LRDs as ``young heavy seeds'' unlike the DCBHs in the MELIORA simulation. This difference in early black hole accretion between \brahma\ and MELIORA could be coming from various aspects of the adopted subgrid prescriptions, including the black hole seeding model, ISM treatment, and stellar and AGN feedback. Disentangling the relative importance of these ingredients is beyond the scope of the present work and will be explored in future studies. Nevertheless, based on their predicted DCBH formation rates and the fraction of seeds undergoing luminous AGN phases, \citet{Cenci_2025} find LRD abundances that are broadly consistent with observations (see their Section~4). While it remains unclear what fraction of these objects would satisfy the observational color-selection criteria adopted in our work, their results demonstrate that at least some fraction of the observed LRD population could plausibly originate from recently formed DCBH seeds. If this channel contributes significantly to the observed LRD population, it would naturally help explain why this \brahma\ realization cannot fully account for the observed LRD abundances. While the more lenient seeding model adopted in this simulation may include some seeds which experience sufficient Lyman-Werner radiation to form via direct collapse, it is not clear that they would appear in similar numbers to a simulation with a Lyman-Werner radiation criterion for black hole seeding. Additionally, neither the simulation nor our mock observation pipeline include any mechanism for these DCBHs to experience the increased accretion rate (and associated brightened AGN) that is often included in young DCBH models of LRDs. %As discussed earlier, if this channel contributes significantly to the observed LRD population, it would naturally help explain why the \brahma\ realization, which does not have this LRD formation pathway, cannot fully account for the observed LRD abundances.

Finally, the redshift evolution of LRD abundances provides an important observational constraint that any theoretical framework for LRD formation must reproduce. Observationally, the LRD abundance increases from $z\sim9$ to $z\sim5$, before declining toward lower redshifts. \citet{Jeon_2025}'s SAM predictions extend to $z\sim5$, and they likewise predict an increasing trend that is broadly consistent with observations over this redshift range. This is perhaps unsurprising given their purely mass-based LRD selection criterion of black holes with $M_{\rm BH}\gtrsim10^6~M_{\odot}$, since such a trend naturally follows from continued black hole seeding and growth. Whether their predicted abundance would also turn over at lower redshifts remains unclear, however, because their analysis does not incorporate the observational LRD color-selection criteria. In our analysis, which does include these criteria, we similarly predict an increasing LRD abundance under the gas-enshrouded AGN SED model from $z\sim8$ to $\sim5$~(compare top and bottom  bottom panels of Figure \ref{fig:LRD_LF}). However, our simulations extend only to $z=5$, preventing us from testing whether the predicted abundance subsequently declines. We also emphasize that our gas-enshrouded SED model represents an upper limit on the LRD abundance, while the true evolution may additionally depend on a redshift-dependent duty cycle of the gas-enshrouded phase. \citet{Cenci_2025} predict that the abundance of recently formed DCBHs declines sharply from $z\sim6$ to $z\sim4$ (their Figure~2), qualitatively resembling the observed decrease in LRD abundances. However, their predicted DCBH formation rate remains largely flat at $z\sim9$ and $z\sim6$, whereas the observed LRD abundance tends to rises over this interval. As in the case of \citet{Jeon_2025}, it remains unclear how the application of the observational color-selection criteria would modify this predicted redshift evolution.

Overall, the comparison between these studies highlights that we are still in the early stages of identifying the physical mechanisms responsible for the LRD population. No single theoretical framework currently reproduces all of the observed properties of LRDs, including their abundances, redshift evolution, black hole masses, and photometric colors. With that being said, the observed population may comprise multiple physically distinct channels, including overmassive black holes undergoing transient gas-enshrouded phases, recently formed DCBHs, super-Eddington accreting Pop III remnants, and compact star-forming galaxies. Disentangling the relative contributions of these channels will require systematic exploration across a broad range of theoretical approaches, from high-resolution simulations that resolve the small-scale physics governing black hole formation and accretion, gas obscuration, and radiative transfer \citep{Inayoshi_2025, Pacucci_2026_DCBH} %Naidu_2025
to large-volume cosmological simulations and semi-analytic models that can predict statistically meaningful LRD populations \citep{Jeon_2025, Cenci_2025, Lumina}.

\section{Conclusions}
\label{sec:Summary}

In this work, we investigate the implications of overmassive black holes for the abundance and properties of objects resembling the JWST-discovered ``little red dots'' (LRDs) within a cosmological framework. To this end, we generate mock observations of AGN and their host galaxies at $z=5{-}8$ using one of the boxes from the \brahma\ simulation suite that produces an overmassive black hole population broadly consistent with current JWST-based BH mass estimates. We model the spectral energy distributions~(SEDs) of both AGN and galaxies, perform synthetic JWST photometry, and apply observational LRD selection criteria to identify simulated analogs. This enables us to characterize the abundance and properties of LRD-like systems in the \brahma\ volume at $z=5$, 6, 7, and 8, spanning the redshift range that contains the majority of current JWST LRD detections.

Despite the presence of overmassive black holes, this \brahma\ realization produces at most two LRDs in four snapshots of its $[36~\mathrm{Mpc}]^3$ volume when analyzed using a standard AGN SED. This corresponds to an LRD abundance at least an order of magnitude below current JWST estimates, demonstrating that the mere presence of overmassive black holes is insufficient to reproduce the observed LRD population.

Motivated by recent studies emphasizing gas-enshrouded AGN as a potential explanation for the Balmer-break-dominated SEDs of LRDs \citep{Setton_2024, Inayoshi_2025, Liu_2025}, we next adopt a gas-enshrouded AGN emission model for the simulated AGN population. This substantially increases the number of LRD analogs identified in the simulation, particularly among galaxies with stellar masses in the range $\rm 10^8 \leq M_{\ast}/M_{\odot} \leq 10^{9.75}$ and black hole masses exceeding $\sim10^7~M_{\odot}$. These systems exhibit bolometric luminosities in the range $\rm 42.9 \leq \log_{10}(L_{bol}/{erg,s^{-1}}) \leq 44.5$, accreting significantly below the Eddington limit. In all simulated LRD analogs, the AGN contributes $\rm \sim 30\%$ or more of the rest-frame optical emission redward of the Balmer break and is primarily responsible for the pronounced red colors that characterize the observed LRD population. Consequently, systems hosting black holes that are luminous relative to their host galaxies are substantially more likely to satisfy current LRD selection criteria.

Under the extreme assumption that all AGN in our simulation snapshots are in a gas-enshrouded phase with an effective duty cycle of unity, we identify \LRDnum\ LRDs, corresponding to an absolute upper limit of $\rm 2.0\times10^{-4}~Mpc^{-3}$ for the predicted LRD number density. But this value does not substantially exceed the current observational estimates, implying that reproducing the observed LRD abundance with the \brahma\ AGN population alone would require near-unity duty cycles for the gas-enshrouded phase. Such high duty cycles are difficult to reconcile with current theoretical expectations \citep{Sun_2026}. Current JWST constraints on the high-z black hole mass function further suggest that this tension may be driven by a deficit of low-mass ($M_{\rm BH}\lesssim10^7~M_\odot$) LRDs in \brahma\ relative to observations, although substantial uncertainties remain in both the inferred black hole masses and the fraction of broad-line AGN classified as LRDs. We conclude that the strongly overmassive black hole population produced in this \brahma\ realization is unlikely, by itself, to account for the observed abundance of LRDs.

At the same time, the absence of an overmassive black hole population appears to be highly detrimental to the formation of LRDs. In the \astrid\ simulation, whose black holes are substantially less massive than those in \brahma\ at fixed stellar mass, the predicted abundance of LRDs remains more than two orders of magnitude below current JWST estimates, even under the extreme assumption of a near-unity duty cycle for the gas-enshrouded phase.

Overall, these findings suggest that while overmassive black holes can substantially enhance the abundance of LRDs in cosmological simulations, their presence alone is insufficient to reproduce the observed abundance of LRDs identified by JWST.

\section*{Acknowledgments}
AKB and PT acknowledge support from NSF-AST 2510738, NSF-AST 2346977, and the NSF-Simons AI Institute for Cosmic Origins which is supported by the National Science Foundation under Cooperative Agreement 2421782 and the Simons Foundation award MPS-AI-00010515.
TDM acknowledges funding from NASA ATP 80NSSC20K0519, NSF PHY-2020295, NASA ATP NNX17AK56G, and NASA ATP 80NSSC18K101, NASA Theory grant 80NSSC22K072.
TDM and YZ acknowledge the support from the NASA FINESST grant NNH24ZDA001N.
FP acknowledges support by the Black Hole Initiative at Harvard University.
LB acknowledges support from NSF award AST-2307171 and NASA award 80NSSC22K0808.
YN acknowledges support from the ITC Postdoctoral Fellowship.
NC acknowledges support from the Schmidt Futures Fund. 
SB acknowledges funding from NASA ATP 80NSSC22K1897 and NSF AST-2509639.
\astrid\ was run on the Frontera facility at the Texas Advanced Computing Center.

%%%%%%%%%%%%%%%%%%%% REFERENCES %%%%%%%%%%%%%%%%%%

% The best way to enter references is to use BibTeX:
\bibliographystyle{mnras}
\bibliography{references} % if your bibtex file is called example.bib

\end{document}